\documentclass[12pt]{article}

\usepackage{scicite}
\usepackage{times}

\usepackage{amsmath}
\usepackage{amsfonts}

\usepackage{booktabs}
\usepackage{dcolumn}

\usepackage{enumitem}

\usepackage{rotating}
\usepackage{graphicx}

\newlist{SubItemList}{itemize}{1}
\setlist[SubItemList]{label={$-$}}

\let\OldItem\item
\newcommand{\SubItemStart}[1]{%
    \let\item\SubItemEnd
    \begin{SubItemList}[resume]%
        \OldItem #1%
}
\newcommand{\SubItemMiddle}[1]{%
    \OldItem #1%
}
\newcommand{\SubItemEnd}[1]{%
    \end{SubItemList}%
    \let\item\OldItem
    \item #1%
}
\newcommand*{\SubItem}[1]{%
    \let\SubItem\SubItemMiddle%
    \SubItemStart{#1}%
}%

 \usepackage{mathtools}
\newcommand{\defeq}{\vcentcolon=}

\newenvironment{sciabstract}{ 
    \begin{quote} \bf }
    {\end{quote}}

\newcommand{\beginsupplement}{%
        \setcounter{table}{0}
        \renewcommand{\thetable}{S\arabic{table}}%
        \setcounter{figure}{0}
        \renewcommand{\thefigure}{S\arabic{figure}}%
     }

\usepackage{graphicx}
\usepackage[position=t,singlelinecheck=off]{subfig}
\usepackage{caption}
\usepackage{float}
\usepackage{multirow}
\usepackage{amsmath}
\usepackage{relsize}

\usepackage[position=t,singlelinecheck=off]{subfig}
\usepackage{caption}

\makeatletter
\DeclareCaptionLabelFormat{mysublabelfmt}{\large\textsf{\Alph{sub\@captype}}}
\makeatother

\newcounter{lastnote}

 \usepackage{color}
 \definecolor{Blue}{rgb}{0,0,1}
 
 \definecolor{Orange}{rgb}{1,0.5,0}
 
 \definecolor{Green}{rgb}{0,1,0}

 \usepackage[position=t,singlelinecheck=off]{subfig}
 \usepackage{caption}
\usepackage{setspace}
\usepackage{float}

\title{\vspace{-13mm}The Wisdom of the Network: How Adaptive Networks Promote Collective Intelligence}

\author
{
    Abdullah Almaatouq$^{1\dagger}$, Alejandro Noriega Campero$^{1\dagger}$, P. M. Krafft$^{2}$,\\ Abdulrahman Alotaibi$^{1}$, Mehdi Moussaid$^{3}$, Alex (Sandy) Pentland$^{1\ast}$ \\ \\
    \normalsize{$^{1}$Massachusetts Institute of Technology, Cambridge, MA, USA}\\
    \normalsize{$^{2}$Oxford Internet Institute, University of Oxford, Oxford, UK}\\
    \normalsize{$^{3}$Max Planck Institute for Human Development, Berlin, Germany}\\
    \normalsize{$^\dagger$ Authors contributed equally to this work}\\
    \normalsize{$^\ast$ To whom correspondence should be addressed; E-mail:  amaatouq@mit.edu}
    \vspace{-5mm}
}

\date{}

\usepackage[british]{babel}
\usepackage{enumitem}

\begin{document} 
    
    
    \baselineskip24pt
    
    
    \maketitle

\begin{sciabstract}    
\setstretch{1.3}{Social networks continuously change as new ties are created and existing ones fade. It is widely noted that our social embedding exerts a strong influence on what information we receive and how we form beliefs and make decisions. However, most empirical studies on the role of social networks in collective intelligence have overlooked the dynamic nature of social networks and its role in fostering adaptive collective intelligence. It has remained unknown (1) how network structures adapt to the attributes of individuals, and (2) whether this adaptation promotes the accuracy of individual and collective decisions. Here, we answer these questions through a series of behavioral experiments and supporting simulations. Our results reveal that social network plasticity, in the presence of feedback, can adapt to biased and changing information environments, and produce collective estimates that are more accurate than their best-performing member. We explore two mechanisms that explain these results: (1) a global adaptation mechanism where the structural connectivity of the network itself changes such that it amplifies the estimates of high-performing members within the group; (2) a local adaptation mechanism where accurate individuals are more resistant to social influence, and therefore their initial belief is weighted in the collective estimate disproportionately. Thereby, our findings substantiate the role of social network plasticity and feedback as key adaptive mechanisms for refining individual and collective judgments.}
\end{sciabstract}

Intelligent systems, both natural and artificial, rely on feedback and the ability to reorganize~\cite{tsypkin1971adaptation, Sosna2019}. Such systems are widespread, and can often be viewed as networks of interacting entities that dynamically evolve. Cell reproduction, for example, relies on protein networks to combine sensory inputs into gene expression choices adapted to environmental conditions~\cite{erwin2009evolution}. Neurons in the brain dynamically rewire in response to environmental tasks to enable human learning~\cite{gutnisky2008adaptive}. Eusocial insects modify their interaction structures in the face of environmental hazards as a strategy for collective resilience~\cite{stroeymeyt2018social}. Fish schools collectively encode information about the perceived predation risk in their environment by changing the structural connectivity of their interaction~\cite{Sosna2019}. In the artificial realm, several machine learning algorithms rely on similar concepts, where dynamically updated networks guided by feedback integrate input signals into useful output~\cite{bottou1998online}. Across the board, the combination of network plasticity (i.e., the ability to reorganize) and environmental feedback (e.g., survival, payoff, reputation, in-sample error) represent a widespread strategy for collective adaptability in the face of environmental changes; providing groups with a practical and easy-to-implement mechanism of encoding information about the external environment~\cite{stroeymeyt2018social, Sosna2019}. 

The emergent ability of interacting human groups to process information about their environment is no exception. People's behavior, opinion formation, and decision-making are deeply rooted in cumulative bodies of social information, accessed through social networks formed by choices of whom we befriend~\cite{wang2010social}, imitate~\cite{zimmermann2004coevolution}, trust~\cite{chow2008social,valenzuela2009there}, and cooperate with~\cite{rand2011dynamic}. Moreover, peer choices are frequently revised, most often based on notions of environmental cues such as success and reliability, or proxies such as reputation, popularity, prestige, and socio-demographics~\cite{kearns2012behavioral, wisdom2013social,henrich2015big,gallo2015effects}.  Human social network ability to reorganize in response to feedback has been shown to promote human cooperation~\cite{rand2011dynamic,gallo2015effects,harrell2018strength} and allows cultural transmission networks over generations to develop technologies above any individual's capabilities~\cite{henrich2015secret, muthukrishna2016innovation}. 

It is widely noted, however, that social influence strongly correlates individuals' judgment in estimation tasks~\cite{muchnik2013social,lorenz2011social,becker2017network,golub2010naive}, compromising the independence assumption (i.e., individual estimate are uncorrelated, or negatively correlated) underlying standard statistical accounts of `wisdom-of-crowds' phenomena~\cite{surowiecki2005wisdom}. 
Additionally, while it is commonly assumed that individuals are correct in mean expectation~\cite{galton1907vox,golub2010naive}, human's independent estimates can be systematically biased~\cite{jayles2017social,indow1977scaling}.
Although the independence and collective unbiasedness assumptions rarely hold in practice, the wisdom of crowds emerges in human groups, nonetheless. In attempts to resolve this puzzle, numerous studies have offered conflicting findings, showing that social interaction can either significantly benefit the group and individual estimates~\cite{bahrami2010optimally,becker2017network,navajas2018aggregated}, or, conversely, lead them astray by inducing social bias, herding, and group-think~\cite{muchnik2013social,golub2010naive,lorenz2011social}. Some notable efforts have focused on providing a partial resolution to inconsistent conclusions. Such studies have found that these divergent effects are moderated by whether well-informed individuals are placed in prominent positions in the network structure~\cite{golub2010naive,becker2017network,moussaid2018dynamical}, how self-confident they are~\cite{bahrami2010optimally, koriat2012two, madirolas2015improving, kearns2009behavioral}, ability to identify experts~\cite{budescu2014identifying}, dispersion of skills~\cite{aral2011diversity,mannes2014wisdom,bonaccio2006advice}, quality of information~\cite{jayles2017social}, diversity of judgments~\cite{davis2014crowd,bonaccio2006advice}, social learning strategies~\cite{barkoczi2016social,toyokawa2019social} and the structure of the task~\cite{toyokawa2019social,mannes2014wisdom}. In other words, whether social interaction is advantageous for the group depends on the environment in which the group is situated. Because people often do not have access to all the parameters of their environment (or the environment can change), it is advantageous to find an easy-to-implement mechanism that performs well across shifting environments.

Theoretical and experimental work on collective intelligence (including the reconciliation efforts mentioned above) has been predominantly limited to frameworks where the communication network structure is exogenous, where agents are randomly placed in static social structures~\textemdash dyads~\cite{bahrami2010optimally, koriat2012two}, fully-connected groups~\cite{lorenz2011social,woolley2010evidence,navajas2018aggregated}, or networks~\cite{golub2010naive,becker2017network}. However, unlike what is explicitly or implicitly assumed in most existing work, the social networks we live in are not random, nor they are imposed by external forces~\cite{kossinets2006empirical}, but emerge shaped by endogenous social processes and gradual evolution within a potentially non-stationary social system. The present study builds on the observation that agent characteristics, such as skill and information access, are not randomly located in network structure. Intuitively, groups can benefit from awarding centrality to---and amplifying the influence of---well-informed individuals. Therefore, the distribution of agents is often the outcome of social heuristics that form and break ties influenced by social and environmental cues~\cite{kearns2012behavioral,wisdom2013social,boyd2011cultural,henrich2015big}, and therefore, the emergent structure cannot be decoupled from the structure of the environment. Hence, we hypothesize that dynamic social influence networks guided by feedback may be central to collective human intelligence, acting as core mechanisms by which groups, which may not initially be wise, evolve into wisdom, adapting to biased and potentially non-stationary information environments.

\section*{Study Design}

\begin{figure}[H]
    \centering
    \includegraphics[width=0.9\columnwidth]{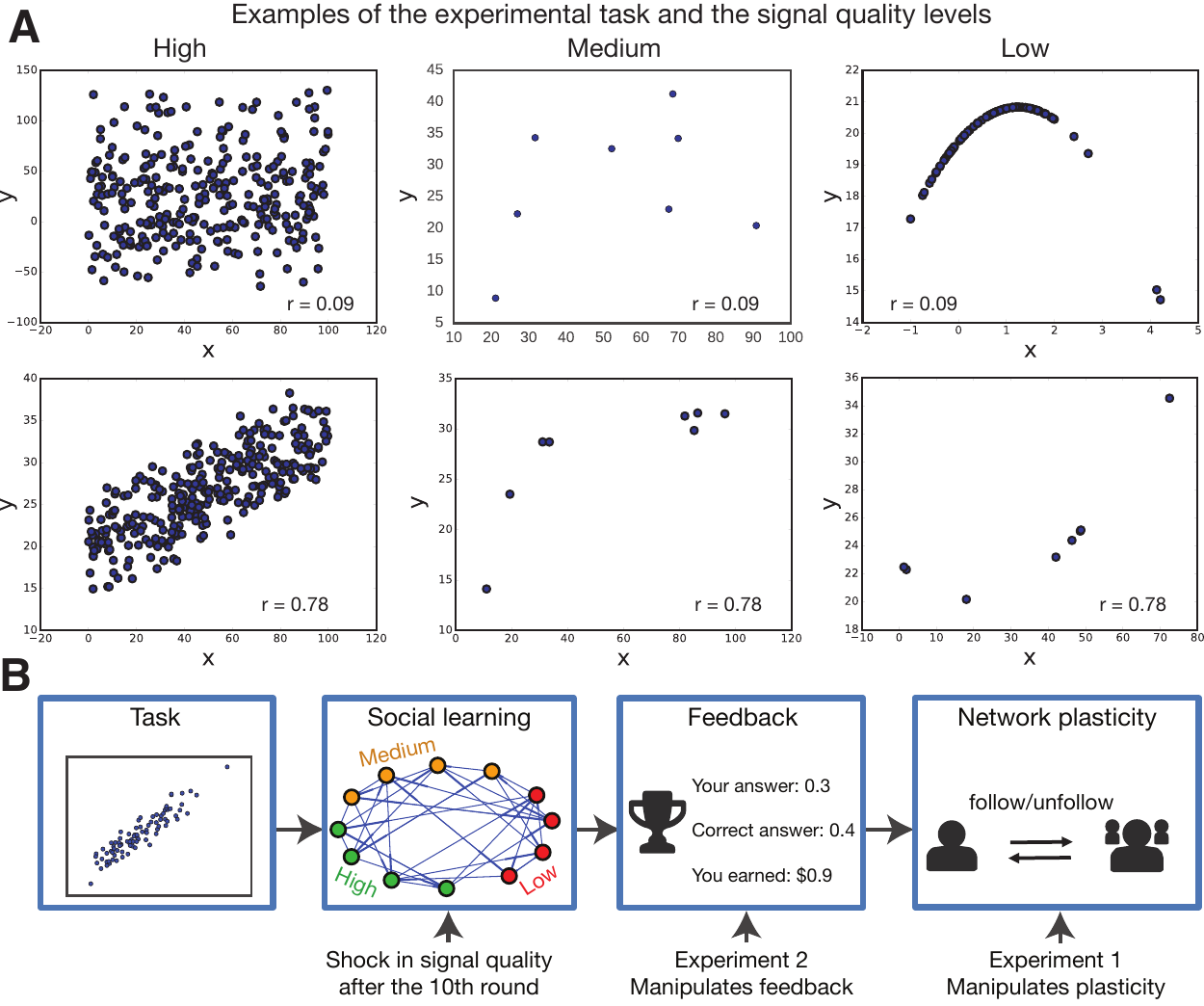}
    \caption{\textbf{Experimental design}. Panel (A) illustrates examples of the scatter plots used in the experiment. For any given round, all participants saw plots that shared an identical true correlation, but signal quality could differ among them. Task signal quality, therefore, could be varied systematically at the individual level by varying the number of points, linearity, and the existence of outliers. Participants were not informed about the signal quality they or other participants were facing. Panel (B) shows an illustration of the experimental design. In experiment 1, the feedback level is fixed (i.e., full-feedback), and network plasticity is manipulated (i.e., static network versus dynamic network). In experiment 2, plasticity is fixed (i.e., always dynamic network), and feedback is manipulated (i.e., no-feedback, self-feedback, and full-feedback). The colors of the nodes represent the signal quality. Each participant experienced a constant signal quality level across the first ten rounds; then, at round eleven, we introduced a shock by reshuffling signal qualities to new levels that stayed constant for the remaining ten rounds.    }
    \label{fig:experiment_design}
\end{figure}

To test the hypothesis that dynamic social influence networks guided by feedback may be central to collective human intelligence, we developed two web-based experiments (i.e., $E_1$ and $E_2$) and a simulation model to identify the role of dynamic networks and feedback in fostering adaptive `wisdom of crowds.' In the two experiments, participants from Amazon Mechanical Turk ($N_{E_1}=719; N_{E_2}=702$) engaged in a sequence of 20 estimation tasks. Each task consisted of estimating the correlation of a scatter plot, and monetary prizes were awarded in proportion to performance at the end of the experiment. Participants were randomly allocated to groups of 12. Each group was randomized to one of three treatment conditions in $E_1$ where we varied the network plasticity or four treatment conditions in $E_2$ where we varied the quality of feedback. Fig.~\ref{fig:experiment_design}A illustrates the overall experimental design. To assess the generality of our findings and tune our intuition, we also simulated a dynamical model of interacting agents in a context similar to our experiments.

\subsection*{Estimation Task: Guess the correlation game} Earlier work demonstrated that estimating the correlation in scatterplots is an intuitive perceptual task that can be leveraged to investigate various aspects of our visual intelligence~\cite{rensink2010perception}. We chose this judgment task for two reasons. First, the task can be carried out rapidly without requiring participants to have specialized skills~\cite{rensink2010perception}. Second, the task structure is simple enough to vary systematically, while still being rich enough to manipulate the quality of the information provided to the participants. In particular, we used scatter plots with three levels of signal quality (varying the number of points and adding outliers or non-linearities; see Fig.~\ref{fig:experiment_design}B;). At every round, all plots seen by participants shared an identical actual correlation, but the quality of the signal could differ among them (Cf.~\cite{moussaid2017reach}). The design also allowed the simulation of a shock to the distribution of information among participants. Specifically, each participant experienced a constant signal quality level across the first ten rounds; then, at round eleven, we introduced a shock by reshuffling signal qualities to new levels that remained constant after that. Participants were not informed about the signal quality they or their peers faced (see SI Section~1 and Figs.~S1-S2 for more details).

\subsection*{Experiment 1 ($E_1$): Varies network plasticity; holds feedback} In the first experiment ($E_1$, $N = 719$), each group was randomized to one of three treatment conditions: a \textit{solo} condition, where each individual solved the sequence of tasks in isolation; a \textit{static network} condition, in which participants were randomly placed in static communication networks; and a \textit{dynamic network} condition, in which participants at each round were allowed to select up to three neighbors to communicate with. Across all conditions, at each round, participants were initially asked to submit an independent guess. Then those in \textit{static} and dynamic network conditions entered a social exposure stage, where they could observe the answers of their network peers, update their own, and see peers' updated beliefs in real-time. After submitting a final guess, participants in all conditions were given performance feedback. Lastly, those in the dynamic network condition were allowed to revise which peers to follow in subsequent rounds (see Fig.~S3 for the experimental design and Figs.~S4-S8 for the online platform screenshots).

\subsection*{Experiment 2 ($E_2$): Varies feedback; holds dynamic network} In the second experiment ($E_2$, $N=702$), each group was randomized to one of four treatment conditions: a \textit{solo} condition, where each individual solved the sequence of tasks in isolation, but this time they were not provided with any performance feedback; a \textit{no-feedback} condition, in which participants placed in a network but were not shown any performance feedback; a \textit{self-feedback} condition, in which participants were placed in a network and shown their own performance feedback; and a \textit{full-feedback} condition, in which participants were placed in a network and shown performance feedback of all participants (including their own). Participants in all conditions in $E_2$ were allowed to revise which peers to follow in subsequent rounds, except for the solo conditions, which acted as our baseline.

\subsection*{Simulation: Varies environmental shock and rewiring rates}
Finally, we simulated interacting agents that update beliefs according to a DeGroot process~\cite{degroot1974reaching}, and rewire social connections according to a performance-based preferential attachment process~\cite{barabasi1999emergence} (see SI Section~2 for model details). Using this model, we explored the effect of plasticity and the quality of feedback to provide further support to our experimental findings, examine the robustness of our findings under different parameter values, and tune our intuition. In these simulations, we also explored the interaction between network adaptation rates\textemdash a network's sensitivity to changes in agents' performance\textemdash and the rate of environmental changes.

\section*{Results}

\subsection*{Individual and collective outcomes}
\begin{figure}[H]
\centering
\includegraphics[width=1\columnwidth]{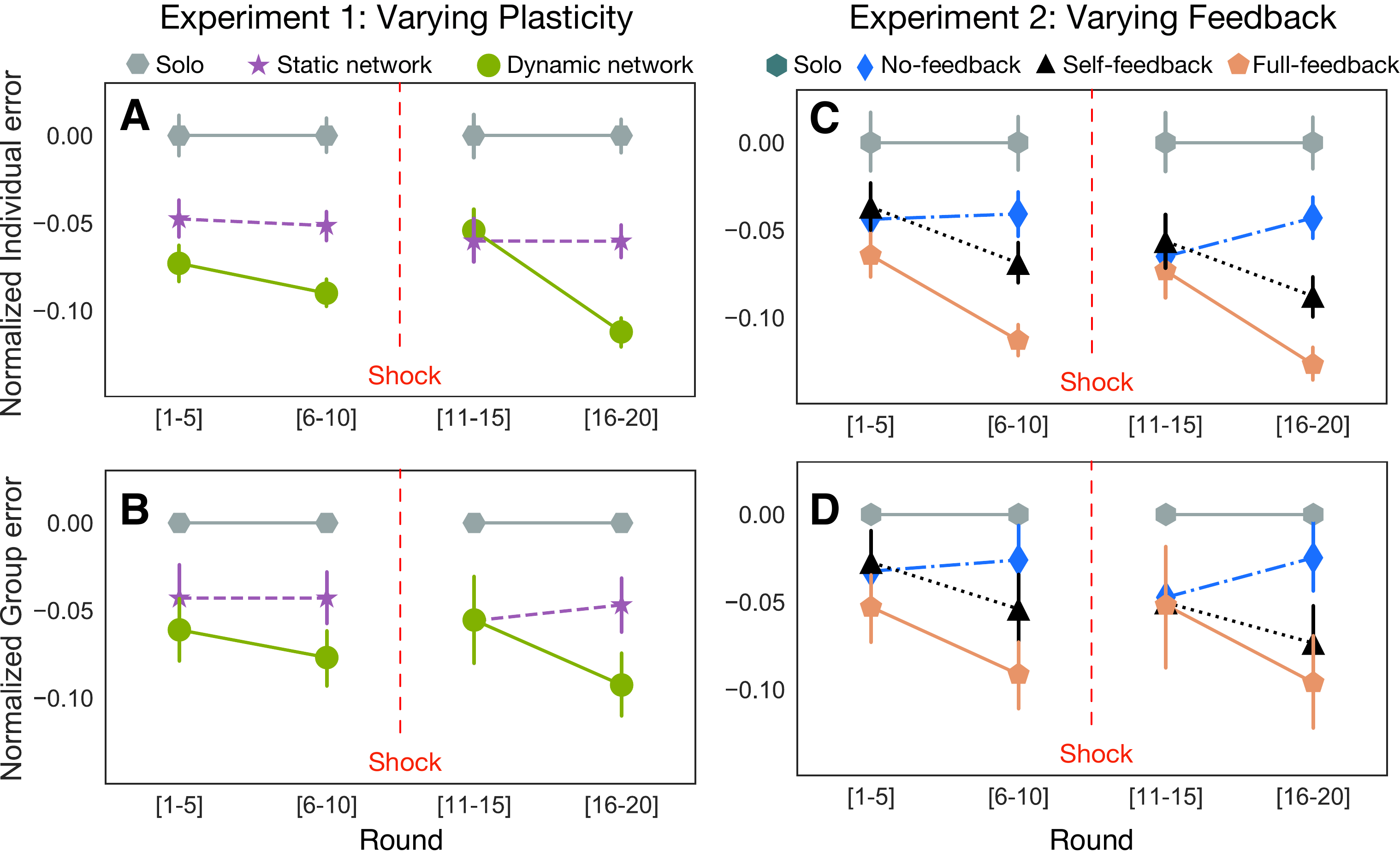}
\caption{\textbf{Individual and collective outcomes}. Groups connected by dynamic influence networks and provided with feedback incur substantially lower individual errors as shown in Panels (A) and (C) and lower collective errors in Panels (B) and (D). The reduction is notably larger and more significant in periods where networks had adapted to the information environment (i.e., rounds $[6,10]$ and $[16,20]$). Errors are normalized with respect to the errors in the \textit{solo} condition. Error bars indicate 95\% confidence intervals.}
\label{fig:errors_per_round}
\end{figure}

We first compared individual- and group-level errors across conditions. Our first result is that networked groups across studies and conditions significantly outperformed equally sized groups of independent participants, which is consistent with prior work on complex tasks~\cite{mason2012collaborative,derex2015foundations} as well as estimation tasks~\cite{becker2017network}. Fig.~\ref{fig:errors_per_round} show the individual and group error rates---using the arithmetic mean as group estimate---normalized with respect to baseline errors in the solo condition of the particular study. Overall, we find that participants in dynamic networks with full-feedback achieved the lowest error rates in both experiments. The dynamic networks, in the presence of feedback, gradually adapted over the course of the experiment. The performance edge was larger in periods where networks had adapted to their information environment (i.e., rounds $[6, 10]\cup [16, 20]$), which we will refer to as the \textit{adapted periods}. 

In particular, in $E_1$ dynamic networks averaged 17\% lower individual error ($\beta =  -0.038$, $z = -4.64$, $P < 10^{-5}$), and 18\% lower group error ($\beta = -0.03$, $z = -3.25$, $P = 0.001$), compared to participants in static networks. In the adapted periods, dynamic networks reduced individual error by 36\% ($\beta =  -0.05$, $z = -6.56$, $P << 10^{-6}$) and group error by 40\% ($\beta =  -0.04$, $z = -4.44$, $P < 10^{-5}$). See Table S1.

Our simulation results corroborate this experimental result. We find that dynamic networks in the presence of feedback adapted to changes in the information environment by shifting influence to agents with better information, substantially decreasing individual and group error over unconnected groups (see Fig.~S10). 

In $E_2$, having self-feedback marginally reduced the overall individual error (7\%; $\beta = -0.015$, $z = -1.38$, $P = 0.17$) and significantly in the adapted periods by (21\%; $\beta = -0.037$, $z = -3.52$, $P = 0.0004$) compared to the no-feedback condition. At the group level, self-feedback averaged 11\% lower error ($\beta = -0.02$, $z = -1.66$, $P = 0.096$) and 29\% in the adapted periods ($\beta = -0.038$, $z = -2.94$, $P = 0.003$).

On the other hand, the full-feedback condition averaged 20\% lower individual error ($\beta = -0.03$, $z = -3.3$, $P = 0.001$), and 16\% lower group error ($\beta = -0.02$, $z = -2.25$, $P = 0.024$), compared to participants in the self-feedback condition. In the adapted periods, full-feedback reduced individual error by 32\% ($\beta =  -0.04$, $z = -4.99$, $P << 10^{-5}$) and group error by 29\% ($\beta = -0.03$, $z = -2.99$, $P = 0.0028$) compared to groups provided with self-feedback (see Table S2).

As the full-feedback condition in $E_2$ and the dynamic network condition in $E_1$ are identical (i.e., both dynamic network and full-feedback), we consider them to be a replication of the same condition across two studies. We confirm that there are no statistically significant differences between the two conditions in Table S3.

In agreement with these experimental findings, simulations confirmed that high-quality feedback is necessary for enabling beneficial group adaptation through social rewiring. Fig.~S11 shows that, as we add more noise to the peer performance feedback, the collective performance of adaptive networks deteriorates until it converges to that of the simple wisdom of crowds (i.e., the solo condition).

\subsection*{The ability of the network to adapt} 
\begin{figure}[H]
\centering
\includegraphics[width=0.95\columnwidth]{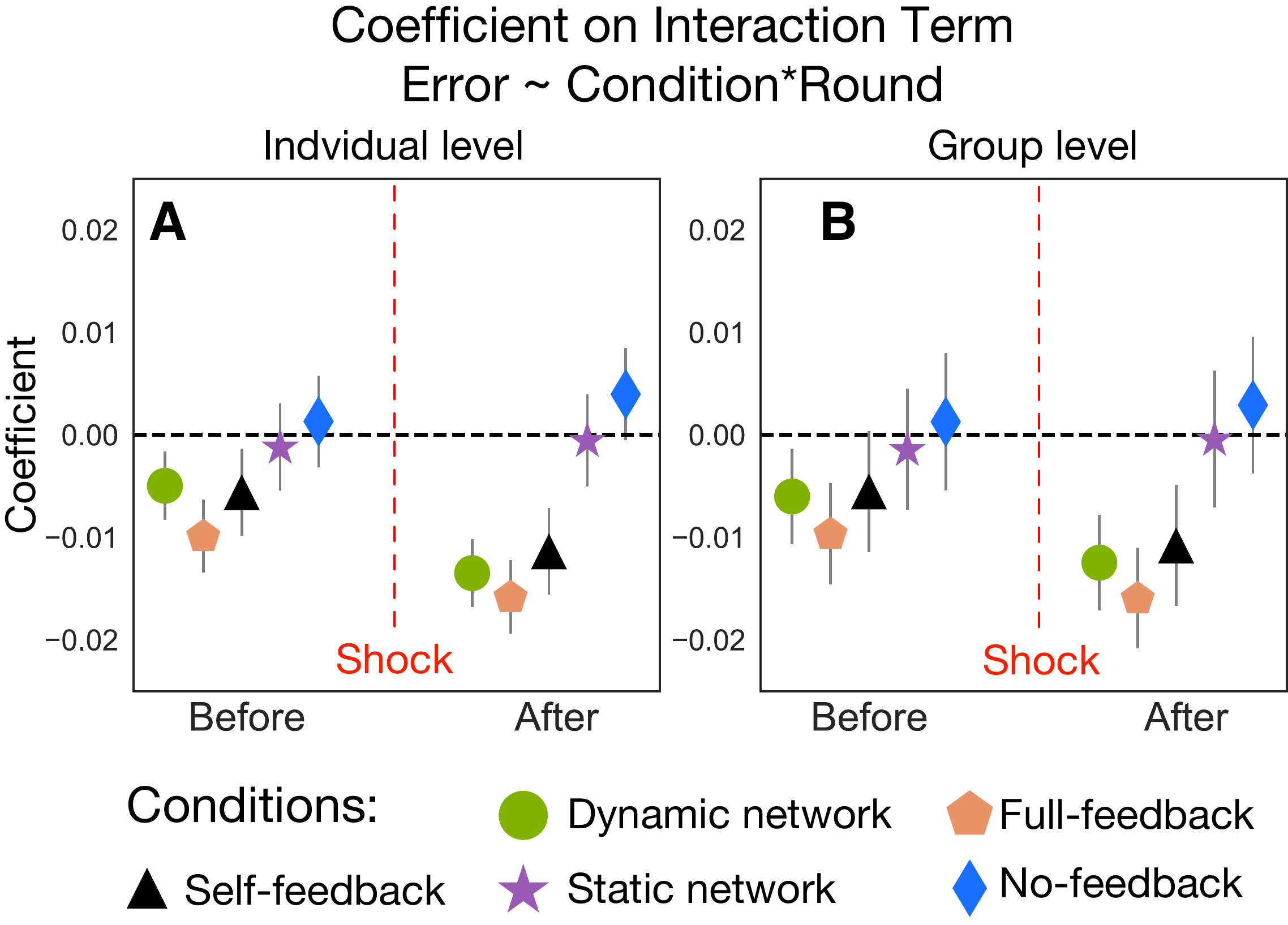}
\caption{\textbf{The ability of the system to adapt}. The coefficients on the interaction term \textit{condition*round} in a mixed effect model to account for the nested structure of the data. Individuals (Panel A) and groups (Panel B) provided with full or partial feedback were able to adapt (i.e., reduce their error as rounds elapsed) both before and after the shock. Error bars indicate 95\% confidence intervals.}
\label{fig:adapt_results}
\end{figure}

We also examined the interaction term between the experimental condition and the experiment round index in a generalized mixed-effects linear model. This allows us to see how the errors across conditions changed as rounds within an experiment elapsed. In particular, a negative coefficient would indicate that individuals or groups are adapting to their environment relative to the unconnected groups, while a positive coefficient would suggest maladaptation---that is, individuals or groups failed to adjust adequately to the environment. Figs.~\ref{fig:adapt_results}A and B show that when both plasticity (i.e., dynamic network) and feedback (full or partial) are provided, the participants could best adapt to the information environment and reduce their error over rounds (i.e., negative coefficients; $P < 0.05$). Only the performance of individuals after the shock in the no-feedback condition was found to be marginally maladaptive  ($\beta = 0.004$, $P=0.084$), while individual and collective performance in the static network condition was neither adaptive nor maladaptive (i.e., flat slope; $\beta = 0; P > 0.4$).

\subsection*{The performance of the best individual} We found that even the best individual did benefit from network interaction. The best individual within each group was determined based on ex-post \textit{revised estimate} performances across all rounds --- that is, based on the quality of the post-social learning estimates. In particular, we find that the best individuals in the dynamic and static conditions from $E_1$ reduced their overall error by roughly 20\% ($P < 10^{-4}$). In $E_2$, the best individuals in the full-feedback condition reduced their error by 30\% ($P < 10^{-4}$), while the best individuals in the self-feedback condition reduced their error by 21\% ($P = 0.009$) and 15\% in the no-feedback condition ($P = 0.057$).

\subsection*{Mean-variance trade-off} 
\begin{figure}[H]
\centering
\includegraphics[width=0.7\columnwidth]{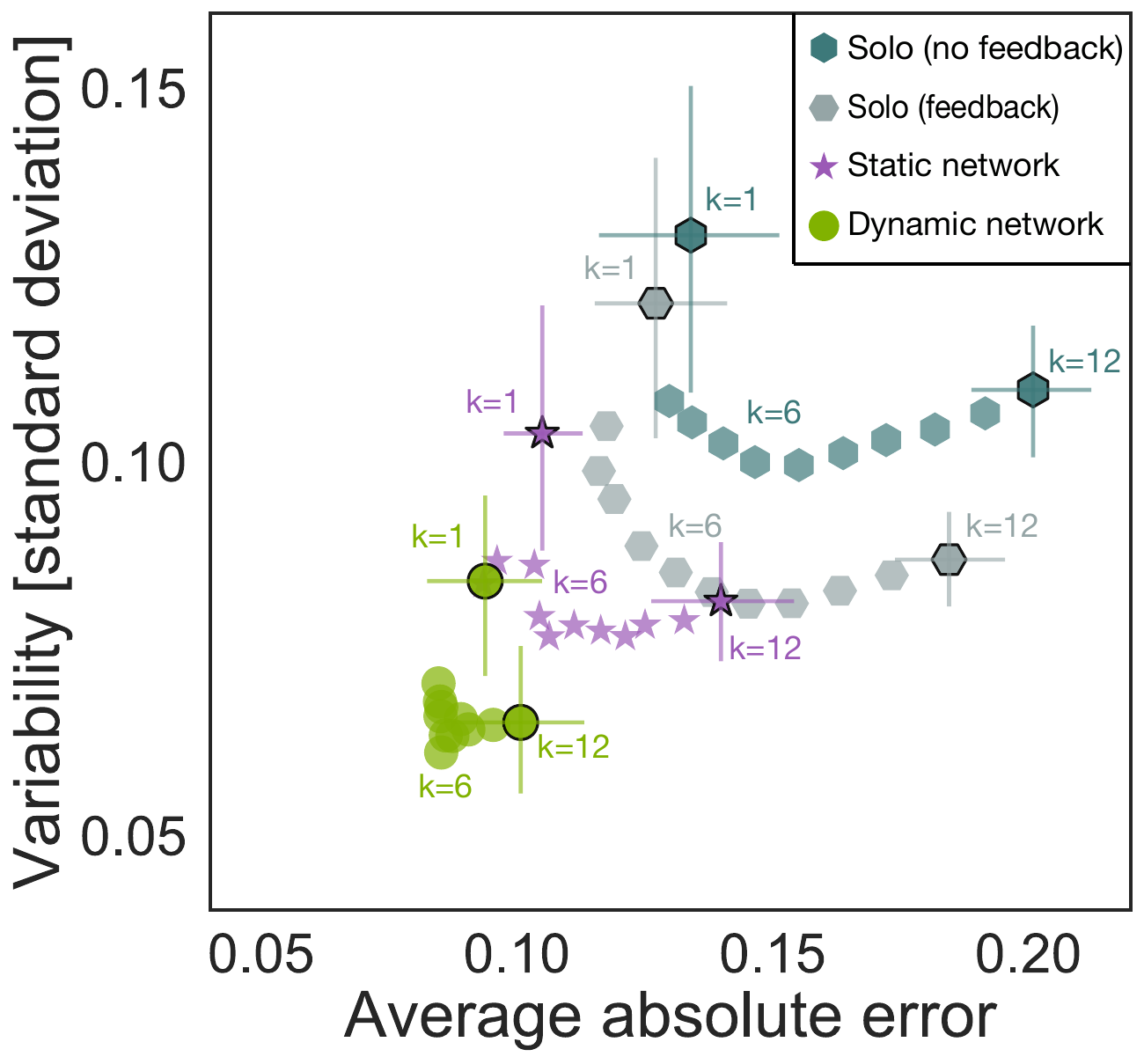}
\caption{\textbf{Mean-variance trade-off.} Mean and standard deviation of absolute errors incurred by \textit{top-k} estimates during the adapted periods. \textit{Top-12} estimates  correspond to the full-group mean, and \textit{top-1} to the group's best individual. Error bars indicate 95\% confidence intervals.
}
\label{fig:trade_off}
\end{figure}

The collective performance of groups was not bounded by that of the best individual. To further examine this, we generalize the use of group means as collective estimates and the definition of \textit{best individual} to analyze the performance of \textit{top-k} estimates\textemdash that is, aggregate estimates where only the guesses of the $k$ best-performing group members are averaged. In particular, \textit{top-12} estimates correspond to the group mean (i.e., the whole-crowd strategy), and \textit{top-1} to the estimates of groups' best-performing individual (i.e., best-member strategy). Fig~\ref{fig:trade_off} reports the mean and standard deviation of estimation errors incurred by the entire range of $k$ (i.e., \textit{top-k} or the select $k$ crowd strategy), estimates during the adapted periods. Ideal estimates would minimize both mean error and variability (i.e., towards the (0,0) corner). The qualitative shape of \textit{top-k} curves reveals that, as we remove low-performing individuals (from $k = 12$ to $k=1$), estimates initially improve in both mean and standard deviation. Then, as we further curate the crowd---roughly---beyond $k = 6$, \textit{top-k} estimates trade-off between decreasing mean error and increasing variability and finally regress in both objectives as $k\rightarrow1$. Roughly speaking, selecting six members strikes a balance between using the estimates of the best members on the one hand and taking advantage of the error-canceling effects of averaging on the other~\cite{mannes2014wisdom}. Interestingly, we find that the full-group average in dynamic networks got 21\% lower error ($P = 0.002$; 500 bootstraps) and 46\% less variability ($P << 10^{-5}$; 500 bootstraps) than the \textit{best individual} in the \textit{solo with feedback} condition (i.e., \textit{dynamic} \textit{top-12} vs. \textit{solo} \textit{top-1}).

\subsection*{Adaptation and environmental shock rates}


Lastly, we explored through simulation the interaction between network adaptation rates\textemdash a network's rewiring sensitivity to changes in agents' performance\textemdash and the arrival rate of environmental shocks. Simulations indicated that networks with higher adaptation rates is suitable for environments with frequent information shocks. Conversely, networks with slower adaptation rates could leverage more extended learning periods, eventually achieving lower error rates in environments with infrequent shocks (see Fig~S10). This short-term versus long-term accuracy trade-off implies that optimal network adaptation rates depend on the pace at which the information environment changes (see Fig.~S12), analogous to notions of optimal adaptation rates in natural systems~\cite{kondoh2003foraging} and learning rates in artificial intelligence algorithms \cite{bottou1998online}.

\subsection*{Adaptive Mechanisms}
The results support our primary hypothesis that network plasticity and feedback provide adaptiveness that can benefit both individual and collective judgment. We explored two social mechanisms to explain these results.

\subsubsection*{Network Centralization}
The first mechanism we identify is a global (or structural) mechanism where dynamic networks in the presence of high-quality feedback adaptively centralized over high-performing individuals. This behavior was predicted by abundant evidence from cognitive science and evolutionary anthropology, which indicates that people naturally engage in selective social learning~\cite{boyd2011cultural,wisdom2013social,henrich2015secret, mannes2014wisdom}---i.e., the use of cues related to peer competence and reliability to choose whom we pay attention to and learn from selectively. Figs.~\ref{fig:network}A and~\ref{fig:network}B show that participants in dynamic networks with full-feedback used peers' past performance information to guide their peer choices. As rounds elapsed, performance information accrued, and social networks evolved from fully distributed into centralized networks that amplified the influence of well-informed individuals. Upon receiving an information shock, the networks slightly decentralized ($\beta=-0.046$, $z=-4.042$, $P < 10^{-4}$; see Table S4), entering a transient exploration stage before finding a configuration adapted to the new distribution of information among participants (Fig.~\ref{fig:network}D for an example of the network evolution). 

\subsubsection*{Confidence Self-Weighting}
\begin{figure}[H]
\centering
\includegraphics[width=1\columnwidth]{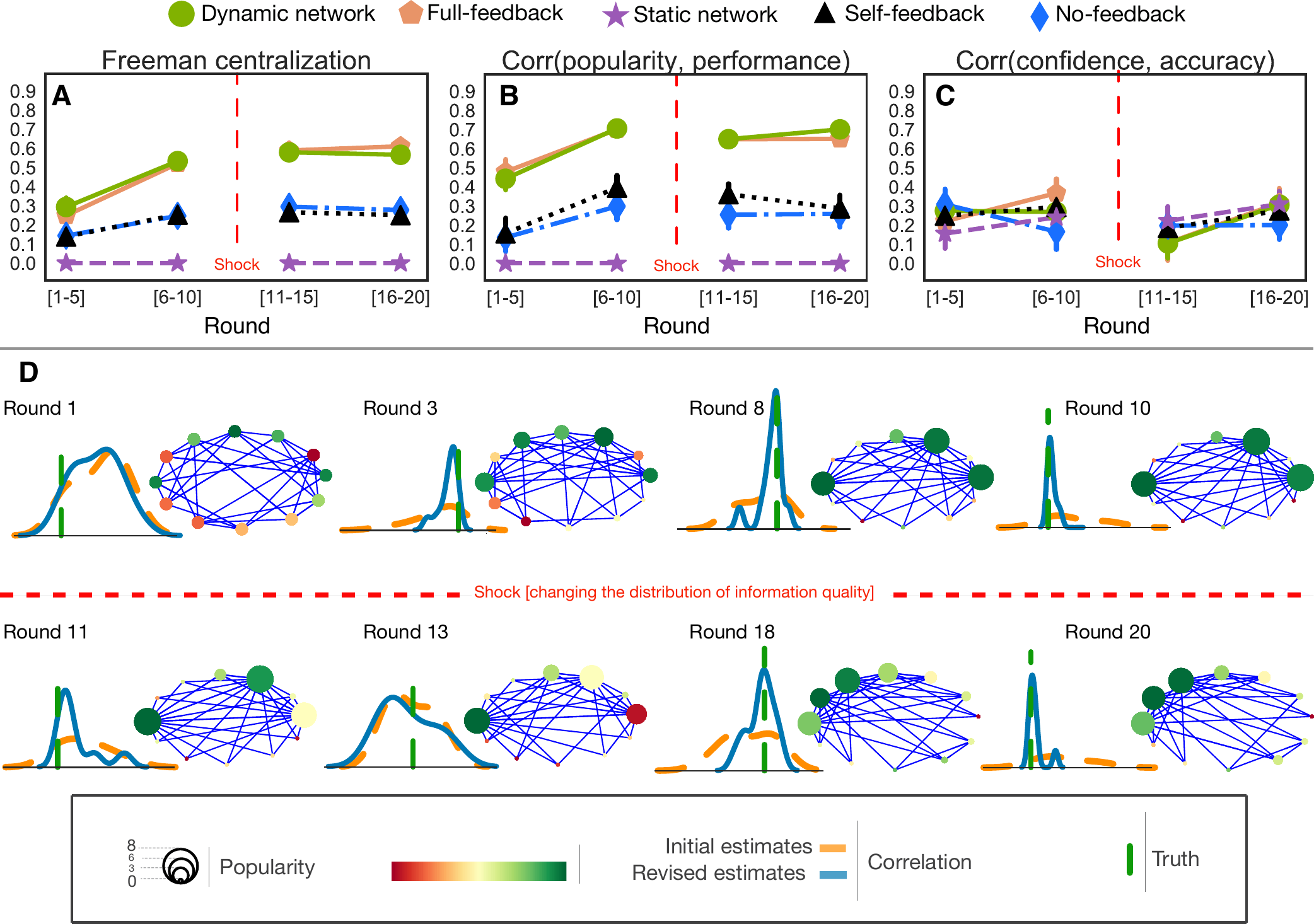} 
\caption{ \textbf{Mechanisms promoting collective intelligence in dynamic networks}.
Panel (A) shows that the network becomes more centralized with time (Freeman global centralization\textemdash, i.e., how far the network is from a 
k). Panel (B) depicts the relation between performance (i.e., average error) and popularity (i.e., number of followers). Panel (C) shows the relationship between the accuracy of the initial estimate and confidence. Error bars indicate 95\% confidence intervals. Panel (D) shows an example of the network evolution in the experiment.}
\label{fig:network}
\end{figure}

Network centralization over high-performing individuals, however, cannot account for all of our results. First, a centralization mechanism alone would suggest that group members may merely follow and copy the best individual among them, hence bounding collective performance by that of the group's top performer, which we found to be untrue (i.e., even the best individual benefits from group interaction). Second, we find that even in the absence of feedback, there is still a correlation between popularity and performance (see Fig.~\ref{fig:network}B) that is similar to the no-feedback condition but weaker than the full feedback condition. Therefore, this alone would not explain why participants in the self-feedback condition in $E_2$ were able to adapt, but not the no-feedback condition.

However, research on the \textit{two-heads-better-than-one} effect indicates that, in the more straightforward case of dyads, even the best individual can benefit from social interaction~\cite{bahrami2010optimally,koriat2012two}; and that the critical mechanism enabling this effect is a positive relationship between individuals' accuracy and their confidence. This is a plausible \textit{local} mechanism that can work in conjunction with both plastic and static networks, as well as with or without feedback. Fig.~\ref{fig:network}C shows that participants in the networked conditions had, overall, a positive correlation between the accuracy of their initial estimates and their self-confidence (measured in terms of resistance to social influence; see Materials and Methods). Participants were likely to rely on private judgments whenever these were accurate and likely to rely on social information otherwise. This positive correlation of confidence and accuracy is consistent with what was found in earlier studies~\cite{becker2017network,madirolas2015improving}. This mechanism allows the participants in the no-feedback conditions to overcome the absence of feedback by exploiting the round-by-round covariation between their own internal decision confidence (i.e., how certain they are) and the answers of their peers~\cite{pescetelli2016perceptual}. That is, if one is confident that they are correct (according to their internal confidence), then they can be equally confident that any neighbor who disagrees with them is wrong, and accordingly down-weight their estimate in future rounds (or break ties with them). Therefore, confidence is used as a signal to learn about the reliability of peers, even in the absence of explicit feedback, and could explain the positive correlation between popularity and performance. Moreover, Fig.~\ref{fig:network}C also shows that, as rounds elapsed, participants in the full- and self- feedback conditions used the explicit performance feedback to calibrate their accuracy-confidence relation further, and were able to re-adapt gradually upon the shock.

\section*{Discussion}

Existing work on collective intelligence and social influence has not considered several aspects that are widespread in natural situations: (1) the rewiring of influence networks; (2) the role of performance feedback; and (3) changing environments (i.e., shocks). We have shown that dynamic influence networks can adapt to biased and non-stationary environments, inducing individual and collective beliefs more accurate than the independent beliefs of the best-performing individual. We also showed that the advantages of adaptive networks are contingent on the presence and quality of performance feedback. 


We acknowledge that the results of laboratory experiments, including ours, rarely translate directly into the real world. Obtaining conclusions of immediate practical relevance would require running a far more extensive and complicated series of experiments than the one we have presented, in which we would vary the available time, group size, task type, group interaction parameters, and many other potentially moderating variables. 

Nonetheless, this study addresses a gap in the literature. Prior work explored the effectiveness of the wisdom of crowds in either unstructured settings (all subjects are exposed to actions of others) or static networks, in which subjects are arranged in non-evolving networks by the experimenter. Additionally, previous work did not study the effect of performance feedback.

The evidence presented here suggests that details of interpersonal communications -- both in terms of the structure of the social interactions and the mechanism of its evolution -- can affect the ability of the system to promote collective intelligence. This is a piece of evidence that dynamism of the networks has profound effects on the processes taking place on them, allowing them to be more efficient and enabling them to adapt to changing environments. 

The insights here provided design guidelines relevant to real-world collective intelligence mechanisms, in contexts such as commodity markets, social trading platforms, crowdfunding, crowd work, prediction markets. We expect the adaptive systems view on collective intelligence to further sprout connections with fields such as social psychology, management science, decision science, evolutionary dynamics, and artificial intelligence, advancing an interdisciplinary understanding and design of social systems and their information affordances.

\section*{Materials and Methods}
\subsection*{IRB, Code, and Data} The study was reviewed and approved by the Committee on the Use of Humans as Experimental participants (COUHES) at MIT. All participants provided explicit consent, and COUHES approved the consent procedure. The experiment was developed using the Empirica platform~\cite{nicolas_paton_2018_1488413}. See OSF repository for experiment implementation, replication data, and code. 

\subsection*{Statistical Tests} All statistics are two-tailed and based on mixed-effect models that included random effects to account for the nested structure of the data. Details of the statistical tests in Tables S1-S4.


\subsection*{Resistance to Social Influence} Our measure of \textit{resistance to social influence} is inspired by the \textit{weight of advice}  (WOA) measure frequently used in the literature on advice-taking~\cite{gino2007effects}. Weight of advice quantifies the degree to which people update their beliefs (e.g., guesses made before seeing the peer's guesses) toward advice they are given  (e.g., the peer's guesses). In the context of our experiments, it is defined as $$WOA \defeq |u_2 - u_1| / |m - u_1|,$$ where $m$ is the neighbors' average initial guess of the correlation, $u_1$ is the participant's initial guess of the correlation before seeing $m$, and $u_2$ is the participant's final guess of the correlation after seeing $m$. Therefore, our measure for \textit{resistance to social influence} is simply $ 1 - WOA$. It is equal to $0$ if the participant's final guess matches the neighbors' average guess, equal to $0.5$ if the participant averages their initial guess and the neighbors' average guess, and equal to $1$ if the participant completely ignores the neighbors' average guess.

\subsection*{Network centralization} We used Freeman centrality~\cite{freeman1978centrality} to quantify the network centralization, which calculates the sum in differences in centrality between the most central node in the network and all other nodes; and then divide this quantity by the theoretically largest such sum of differences (a star network of the same size):
$$
C_{x} ={\frac  {\sum _{{i=1}}^{{N}}C_{x}(p_{*})-C_{x}(p_{i})}{\max \sum _{{i=1}}^{{N}}C_{x}(p_{*})-C_{x}(p_{i})}},
$$ where $C_x(p_i)$ is the in-degree (number of followers) of individual $i$, $C_x(p_*)$ is the in-degree of the most popular individual, and  $\max \sum _{{i=1}}^{{N}}C_{x}(p_{*})-C_{x}(p_{i}) = \left [ (N-1) (N-2) \right]$ is the theoretically largest sum of differences. 

\bibliographystyle{Science}
\bibliography{references}

\newpage

\section*{Supplementary Materials}

\beginsupplement

\subsection*{Access to data and code}
All of the data and analysis code are publicly available at the Open Science Framework (OSF) repository. The study was reviewed and approved by the Committee on the Use of Humans as Experimental participants (COUHES) at MIT. All participants provided explicit consent to participants in this study, and COUHES approved the consent procedure. The experiment was developed using the Empirica (https://empirica.ly/) platform, an open-source ``virtual lab'' framework, and a platform for running multiplayer interactive experiments and games in the browser~\cite{nicolas_paton_2018_1488413}.

\section{Details of Experimental Setup}
\subsection*{Guess the Correlation Task}
\begin{figure}[H]
    \centering
    \includegraphics[width=0.95\textwidth]{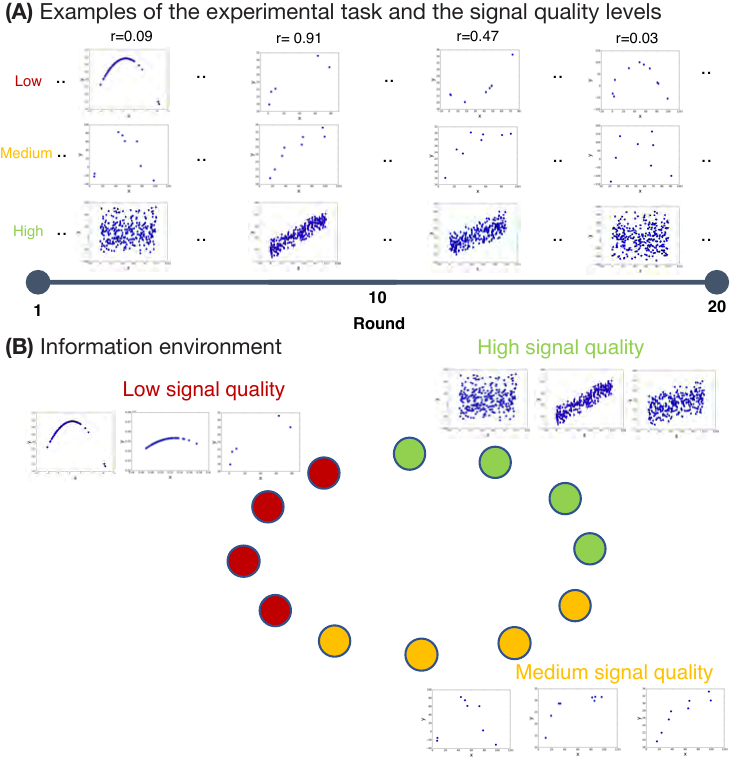}
    \caption{{Guess the Correlation Game}. An illustrative example of the scatter plots used in the experiment is shown in Panel (A). The quality of the signal, therefore, could be varied systematically at the individual level by varying the number of points, linearity, and the existence of outliers. All participants saw plots that shared an identical true correlation, but signal quality levels could differ among them, as shown in Panel (B). Participants were not informed about the signal quality level they or other participants were facing.    }
    \label{fig:task_examples}
\end{figure}

\begin{figure}[H]
    \centering
    \includegraphics[width=1\columnwidth]{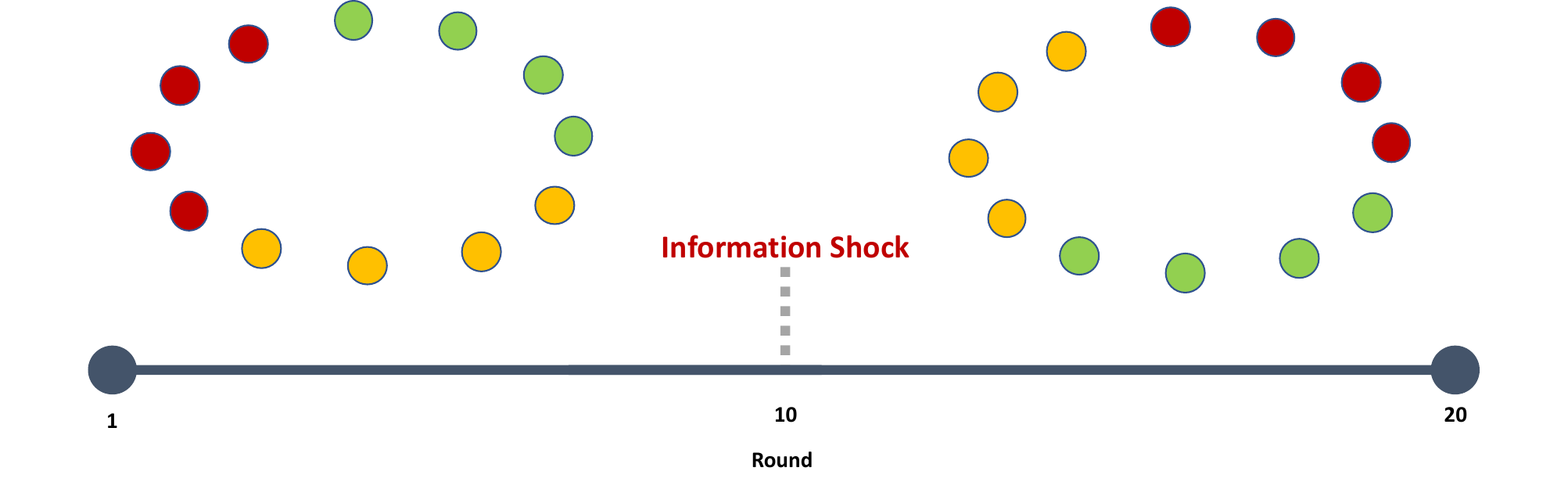}
    \caption{{Shock to the Information Environment}. We provide a change in the environment after round 10 by changing the signal quality levels for the participants for the remainder of the experiment and thereby we simulate non-stationary distributions of information among participants.
    }
    \label{fig:shock}
\end{figure}

\begin{figure}[H]
\includegraphics[width=1\textwidth]{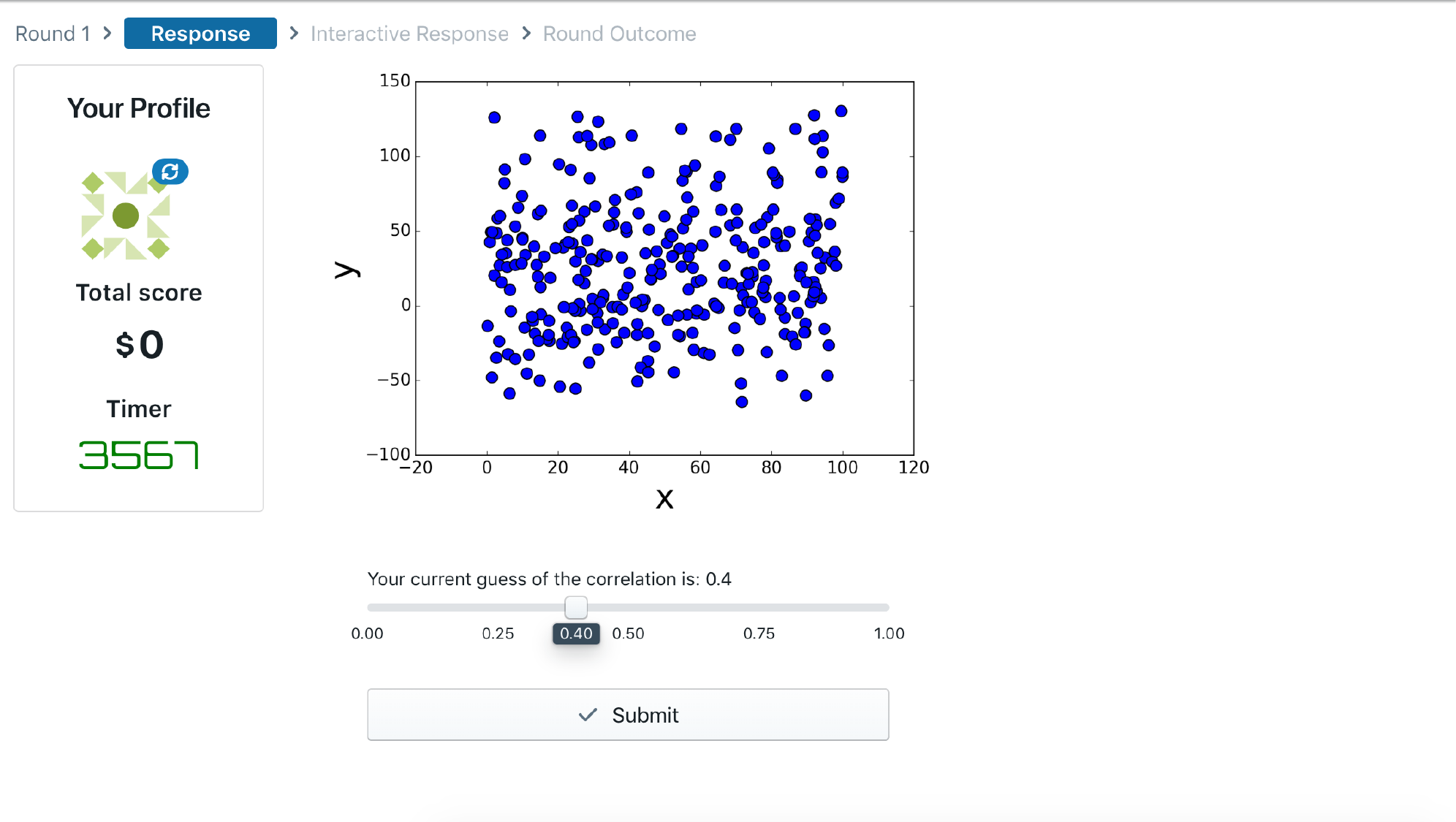}
\caption{Participants in all conditions make independent guesses about the correlation of two variables independently.}
\label{won:screenshot1}
\end{figure}

\begin{figure}[H]
\includegraphics[width=1\textwidth]{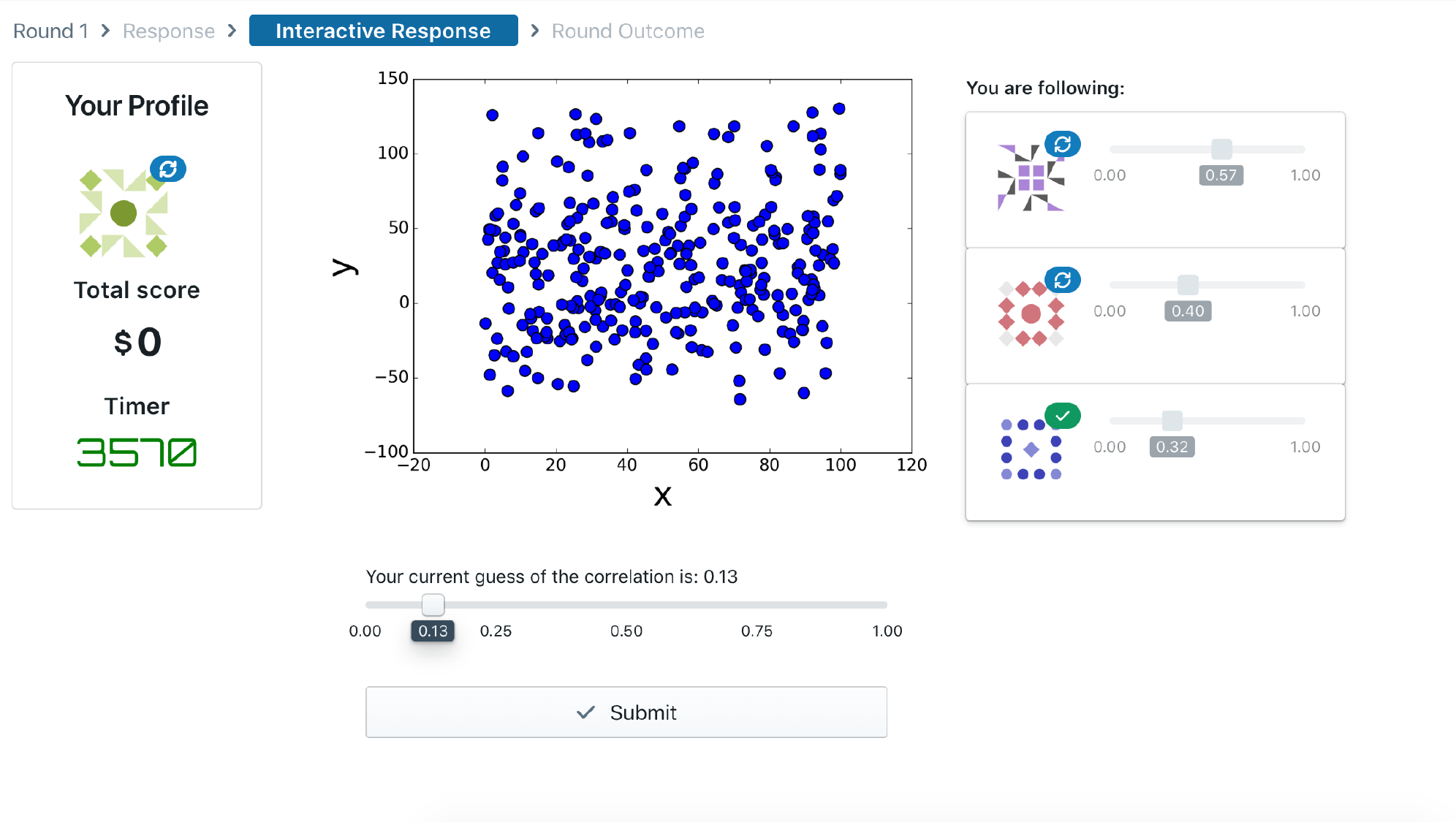}
\caption{Participants in the network condition engage in a an active social learning phase, where they are exposed to their ego-network's estimates in real time.}
\label{won:screenshot2}
\end{figure}

\begin{figure}[H]
\includegraphics[width=1\textwidth]{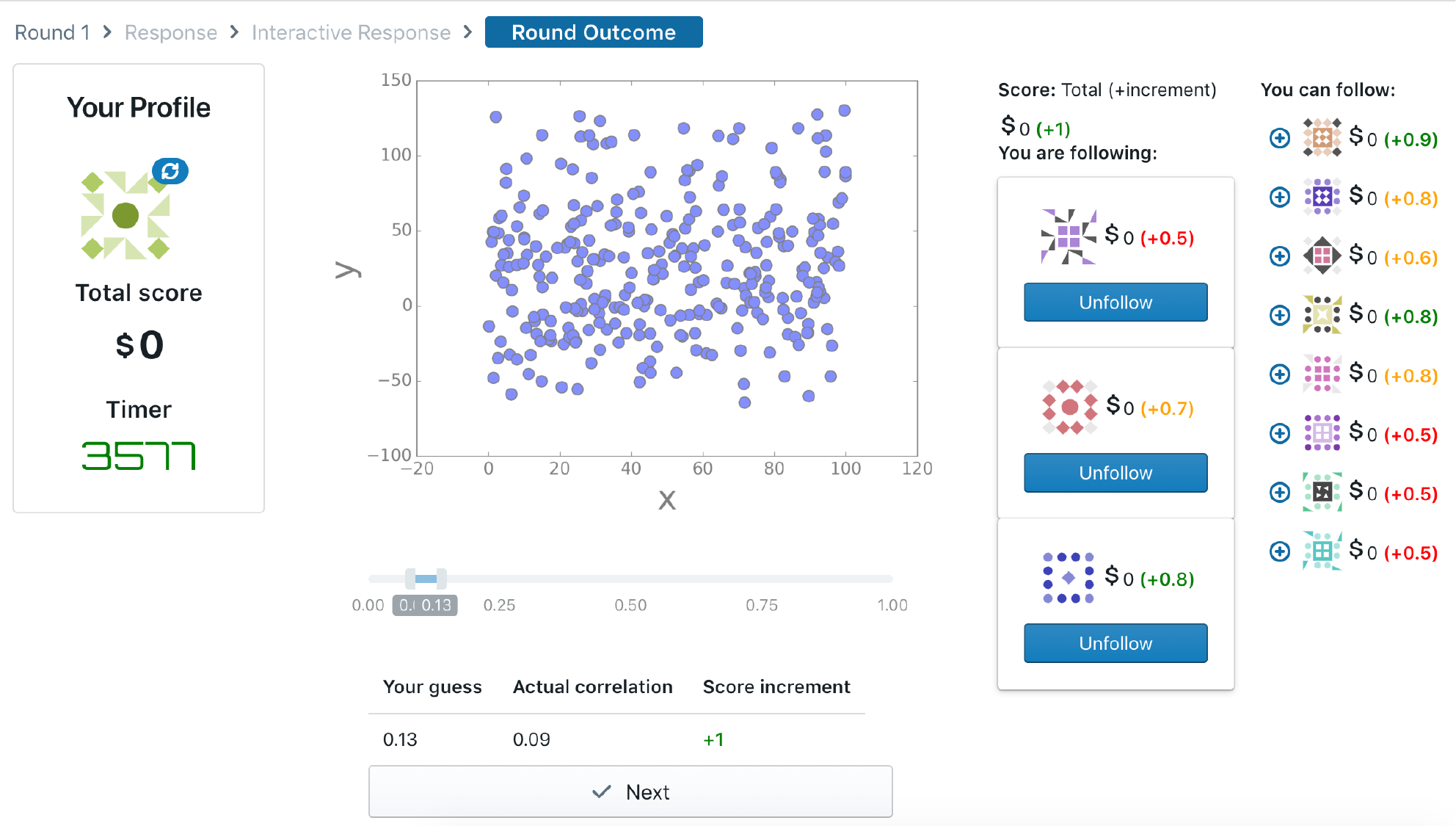}
\caption{After each task round, participants in the feedback conditions see the appropriate level of feedback for the conditions. This figure illustrates the dynamic network condition with full feedback (i.e., as opposed to no-feedback or only self-feedback). In all of our experiments, the maximum number of outgoing connections is three.}
\label{won:screenshot3}
\end{figure}

Participants were prompted to estimate the correlation from a scatter plot (namely, ``Guess the correlation'' game) and were awarded a monetary prize based on the accuracy of their final estimate. This estimation task is designed to expose the mechanisms that allow intelligent systems to adapt to changes in their information environment. We can influence the performance level of participants by implementing three signal quality levels (e.g., adjusting the number of points or linearity): high, medium, and low. At every round, all plots seen by participants shared an identical true correlation, but signal quality levels could differ among them (see Fig.~\ref{fig:task_examples}). The allowed us to construct the environment and provided us with the ability to simulate a shock to the distribution of information among participants. Specifically, each participant experienced a constant signal quality level across the first ten rounds; then, at round eleven, we introduced shocks by reshuffling signal qualities to new levels that remained constant after that (see Fig.~\ref{fig:shock}). Participants were not informed about the signal quality levels they or their peers faced. For a screenshot of the experiment, see Fig.~\ref{won:screenshot1}, Fig.~\ref{won:screenshot2}, and Fig.~\ref{won:screenshot3}.

\subsection*{Experiment 1: Manipulates network plasticity; full feedback}
\begin{figure}[H]
    \centering
    \includegraphics[width=1\columnwidth]{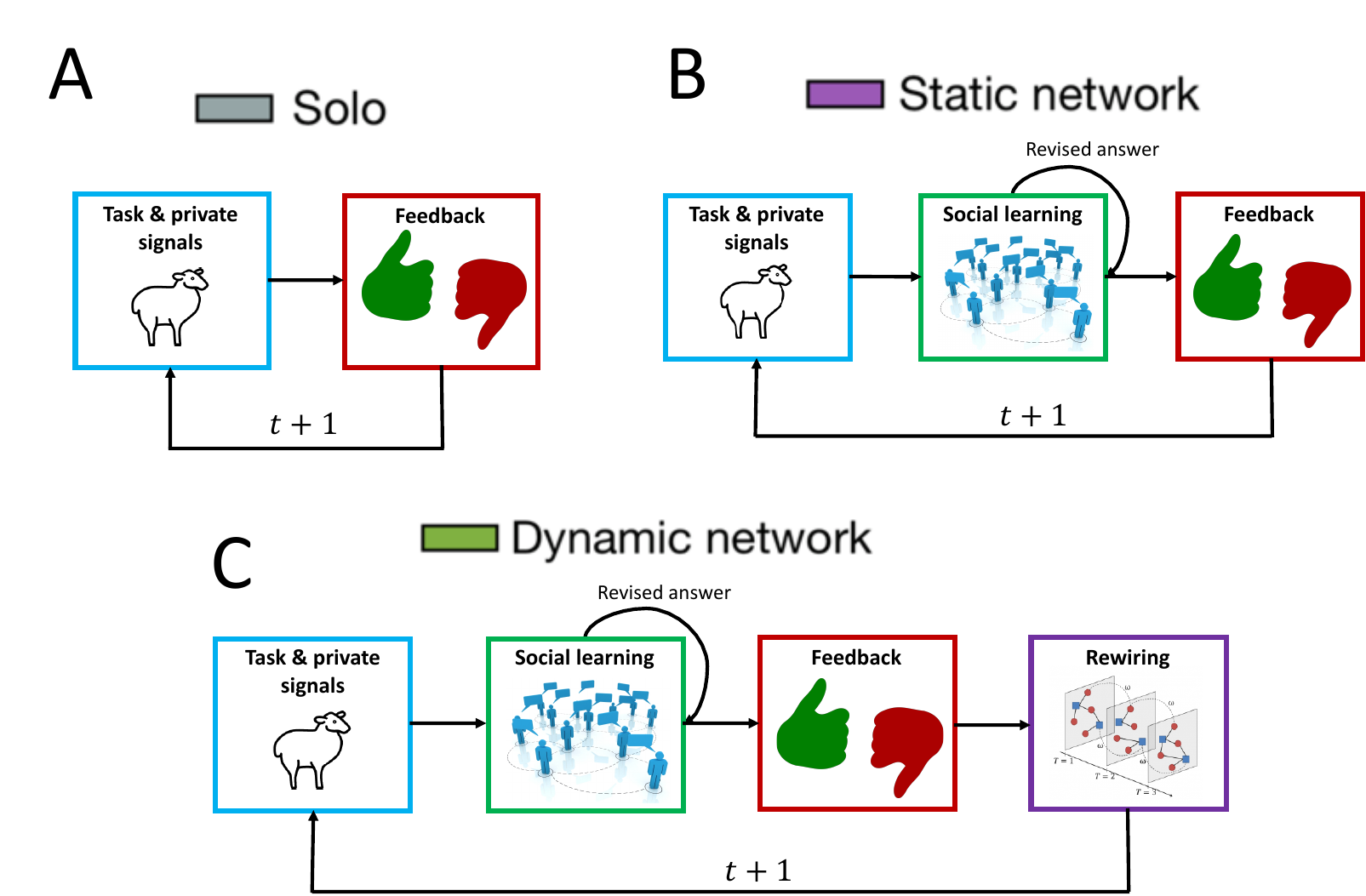}
    \caption{{Illustration of the experimental conditions in study 1.} Panel (A) depicts the Solo condition (i.e., no social information) where participants make independent estimates. This condition corresponds to the baseline wisdom of the crowd context. Panel (B) describes the Static network condition (i.e., social learning) where participants engage in a stage of interactive social learning, where they are exposed to the estimates of a fixed set of peers in real-time. Panel (C) describes the Dynamic network (i.e., selective social learning) condition that adds the possibility for participants to choose who to follow and be influenced by in the next round.}
    \label{fig:WON_exp1}
\end{figure}

In $E_1$, each group was randomized to one of three treatment conditions:
\begin{itemize}
    \item \textit{solo} condition: each participant solved the sequence of tasks in isolation (i.e., no social information). This condition corresponds to the traditional `wisdom of the crowds' context~\cite{surowiecki2005wisdom,prelec2017solution}. See Figure~\ref{fig:WON_exp1}A.
    \item \textit{static} condition: participants were randomly placed in static communication networks. That means participants will engage in a stage of active social learning, where they are exposed to their ego-network's estimates in real-time. See Figure~\ref{fig:WON_exp1}B. This context is analogous to that studied by work at the intersection of the `wisdom of crowds' and social learning, such as as~\cite{lorenz2011social,golub2010naive}.
    \item \textit{dynamic} condition:  participants at each round were allowed to select up to three peers to follow (i.e., get the ability to communicate with) in subsequent rounds. See Figure~\ref{fig:WON_exp1}C. This condition is novel to the work of this dissertation.
\end{itemize}

Note that in the \textit{social learning} stage (i.e., in the static and dynamic conditions; see Figure~\ref{fig:WON_exp1}), participants observe in real-time the estimates of the other participants that they are connected to and can update their estimates multiple times before they submit their final estimate. It is up to the participant to decide how to update their guess to accommodate the information and experiences, the opinions and judgments, the stubbornness and confidence, of the other players. After submitting a final estimate, participants in all conditions were given performance feedback. That included how much they earned, what was the correct correlation, what was their guess.

\subsection*{Experiment 2: Manipulates feedback; dynamic network}
In $E_2$, each group was randomized to one of four treatment conditions:

\begin{itemize}
    \item \textit{solo} condition, where each individual solved the sequence of tasks in isolation.
    \item \textit{no feedback} condition, in which participants were not shown performance feedback.
    \item \textit{self feedback} condition, in which participants were shown their performance feedback.
    \item \textit{full feedback} condition, in which participants were shown scores of all participants (including their own)
\end{itemize}

Participants in all conditions (except solo, our baseline) were allowed to revise which peers to follow in subsequent rounds (i.e., similar to the `dynamic network` condition in study 1).

\subsection*{Participant Recruitment}
All participants were recruited on MTurk by posting a HIT for the experiment, entitled ``Guess the correlation and win up to \$10'', a neutral title that was accurate without disclosing the purpose of the experiment. All participants provided explicit consent to participants in this study, and COUHES approved the consent procedure. All data collected in the experiment could be associated only with the participant's Amazon Worker ID on MTurk, not with any personally-identifiable information. All players remain anonymous for the entire study. At the beginning of a session, participants read on-screen instructions for the condition they are randomly assigned. Participants could start the experiment only once they have completed a set of comprehension questions.

\section{Numerical Simulations}
We implemented numerical simulations where we focus on two conditions: (1)  traditional wisdom of crowds (i.e., independent actors with individual feedback); and (2) adaptive wisdom of crowds (i.e., dynamic networks with full feedback). To follow the properties of our framework, we operationalized social learning as a DeGroot process~\cite{degroot1974reaching}, and propose a performance-based preferential detachment and attachment model for the network rewiring heuristics.

\textbf{Notation}. Let $N = \{1,2,...,n\}$ represent a group of agents that participate in a sequence of tasks, indexed by discrete time $t$. Let $G(N, E^{(t)})$ be a sequence of directed graphs representing the influence network at each period $t$. Let $e_{ij}^{(t)}\in[0,1]$ denote the edge weight of $(i,j)$ at time $t$, and $M^{(t)}$ the row-normalized stochastic matrix associated with $E^{(t)}$, i.e.,  $M_{ij}^{(t)} = \frac{e_{ij}^{(t)} } { \sum_{ h \in N} e_{ih}^{(t)} }$.

Agents receive private signals $s_i^{(t)}\in [0,1]$, for $i\in N$, regarding the true state of the world $\omega^{(t)}\in [0,1]$. Similarly, we denote agents' post-social learning beliefs by $p_i^{(t)}\in [0,1]$, for $i\in N$. 

\begin{figure}[H]
\centering
\includegraphics[width= 1\columnwidth]{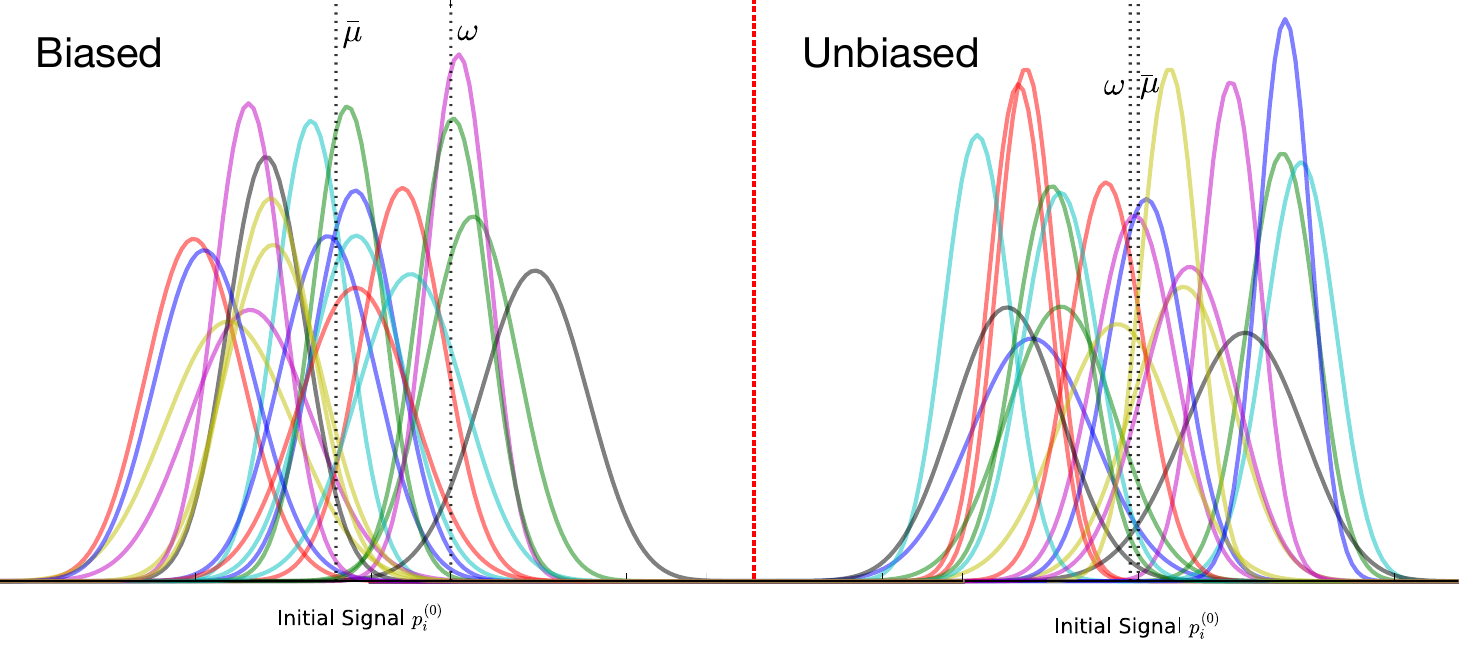}
\caption{Traditional accounts of `wisdom of crowds' phenomena assume unbiased and statistically independent signals among agents. In our model, we assume arbitrary (potentially biased) initial signals.}
\label{fig:si_signals}
\end{figure}

\textbf{Private Signals}. We depart from the commonly made assumption of collective unbiasedness of agents' private signals \cite{surowiecki2005wisdom,lorenz2011social,demarzo2001persuasion,golub2010naive}, allowing agents' signals to be distributed with arbitrary means and skewness. Let $\mu_i = E[s_i]$ denote the mean of agent $i$'s signal, and $\bar{\mu} = \frac{1}{n}\sum_{i} \mu_i$ be the collective mean of private signals; we are interested on the more general setting of information environments where $\bar{\mu} \neq \omega$, i.e., where the collective distribution of initial signals is not centered on the truth. Figure~\ref{fig:si_signals} illustrates the difference between unbiased and biased information environments.

\textbf{Social Learning Process.} Social learning is modeled as a DeGroot process~\cite{degroot1974reaching}, where each agent updates her belief by taking weighted averages of her own belief (i.e., private signal) and the beliefs of neighboring agents. Averaging as social learning heuristic has been well studied empirically and theoretically \cite{demarzo2001persuasion,golub2010naive}, and shown to robustly describe real-world belief updating better than more optimal rational Bayesian models~\cite{chandrasekhar2012testing}. 
In particular, we model post-social learning beliefs as the result of a two-stage DeGroot process on private signals, given by
\begin{equation}
p^{(t)} =  \big( {M^{(t)}} \big)^2  \hspace{1pt} s^{(t)}    
 \label{eq:degroot-2}
\end{equation}

\textbf{Individual performance} is evaluated based on the errors of post-social influence estimates. Individual cumulative error is defined by:
\begin{equation*}
\epsilon_{i}^{(t)}  
= \hspace{4pt}\frac{1}{\lambda+1} \sum_{r\in[0,\lambda]}
\Big| p_i^{(t-r)}  - \omega^{(t-r)} \Big|,
\label{eq:performance-lambda}
\end{equation*} where $\lambda$ controls the number of retrospective periods that performance information is averaged across.  

Agents assess performance of other agents relative to the performance of the best agent in the group. We define relative error of agent $i$ as $\pi_i^{(t)} =  \epsilon_i^{(t)} \hspace{1mm} -\epsilon_{min}^{(t)} $ , and denote the set of performance information available to agent $i$ at time $t$ by vector $\Pi_i^{(t)} \in [0,1]^n$ with elements 
$$
\pi_{ij}^{(t)} = \hspace{4pt} 
  \begin{cases} 
      \hfill  \hspace{4pt}  \pi_{j}^{(t)}   \hfill    & \text{ for $j \neq i$} \vspace{1mm} \\
      \hfill  \hspace{4pt}  \pi_{s_i}^{(t)}   \hfill  & \text{ for $j = i$} \\
  \end{cases}
$$
where $\pi_{s_i}^{(t)}$ is the relative error of agent $i$'s private signal.

\textbf{Collective error}. We are interested on the wisdom of the dynamic network (WDN) error, $\epsilon_{wdn}$, which captures collective error after selective social learning according to the interaction network. We compare $\epsilon_{wdn}$ against the wisdom of the crowd (WC) baseline, $\epsilon_{wc}$, which captures collective error of the simple averaging of agents' initial signals.
\begin{equation}
\epsilon_{wdn}^{(t)} = \big|  \omega^{(t)} - \frac{1}{n}\sum_{i} p_i^{(t)}   \big|
\label{eq:pi_wdn}
\end{equation}
\begin{equation}
\epsilon_{wc}^{(t)} = \big|  \omega^{(t)} - \frac{1}{n}\sum_{i} s_i^{(t)}   \big|
\label{eq:pi_wc}
\end{equation}

\textbf{Influence Rewiring Process}. Individuals connect by weighted influence links that are revised over time. We model influence rewiring heuristics that strengthen links when a neighbor exhibits high performance and weaken or break links when a neighbor performs poorly. Agents can distribute attention among a limited number of peers,  captured by parameter $\kappa$, which represents cognitive or infrastructure constraints (e.g., limits on our ability to keep track of social information and relations~\cite{dunbar1998social}). In particular, agents dynamically allocate $\kappa \in N$ shares of their attention to other agents. Let $e_{ijk} \in \{0,1\}$, for $k \in \{1,2,...,\kappa \}$, indicate that $i$ places attention share $k$ on $j$, then $e_{ij} = \sum_{ k} e_{ijk}$ and $e_{ij} \in \{0, 1, 2,..., \kappa \}$.

\textbf{Probability of detachment}. Probability that agent $i$ detaches from $j$ is a positive function of $i$ and $j$'s errors, and given by equation \ref{eq:beta}. For example, if $i$'s error is among the lowest of the group ($\pi_i^{(t)} \approx 0$), $i$ is unlikely to rewire her local network. Conversely, if $i$'s error is significant (e.g., $\pi_i^{(t)} \approx 1$), $i$ detaches from $j$ with probability dependent on $j$'s error. 
\vspace{-3mm}
\begin{equation}
\beta_{ij}^{(t)} \hspace{4pt} = \hspace{4pt} 
      \hfill \Big( \pi_{i}^{(t)} \hspace{3pt}  \pi_{ij}^{(t)} \Big) ^{\frac{1}{2}}  \label{eq:beta}
\end{equation}

\textbf{Probability of Attachment}. High-performing agents are more likely to be followed. Analogous to generalized preferential attachment \cite{barabasi1999emergence}, probability that agent $i$ attaches to $j$ is inversely related to $j$'s error, and given by 
\vspace{-3mm}
\begin{equation}
\alpha_{ij}^{(t)} \hspace{4pt} = \hspace{4pt} 
  \Bigg( \frac{1 - \pi_{ij}^{(t)}}{n - \sum\limits_{ j} \hspace{0pt} \pi_{ij}^{(t)}}    \Bigg)^{2} c
\label{eq:alpha}
\end{equation} 
where $c$ is a normalization constant.

\textbf{Network Evolution}. Define i.i.d. random variables $b_{ijk}^{(t)}$  \textit{$\sim$ Bernoulli$\big(\beta_{ij}^{(t)}\big) \hspace{6pt} \forall (i,j,k)$}, then random variables $b_{ij}^{(t)} = \sum_{ k} b_{ijk}^{(t)} e_{ijk}^{(t)}$  \textit{$\sim$ Binomial$\big( e_{ij}^{(t)}, \beta_{ij}^{(t)}\big)$ } indicate the amount of attention shares that $i$ detaches from $j$ in period $t$. Define $n$-dimensional random vectors $a_{i}^{(t)}$ $\sim$ \textit{Multinomial}$\big( \hspace{2pt} \sum_{ j} b_{ij}^{(t)} \hspace{4pt} , \hspace{4pt} \alpha_{i}^{(t)} \hspace{2pt} \big)$ , where $\alpha_{i}^{(t)}$ is $i$'s vector of attachment probabilities. Elements $a_{ij}^{(t)} \in \{0,1,...,\kappa\}$ indicate the amount of shares that $i$ attaches to $j$ in period $t$ , and network evolution is given by
\vspace{-3mm}
\begin{equation}
e_{ij}^{(t+1)} = \hspace{4pt} e_{ij}^{(t)} \hspace{2pt} - b_{ij}^{(t)} \hspace{2pt} + a_{ij}^{(t)}    
\hspace{55pt} \forall i,j
\label{eq:net-evolution}
\vspace{-4mm}
\end{equation} 
\vspace{-5mm}

\section{Supplementary Figures}
\vspace{-2mm}

\begin{figure}[H]
\includegraphics[width=1\textwidth]{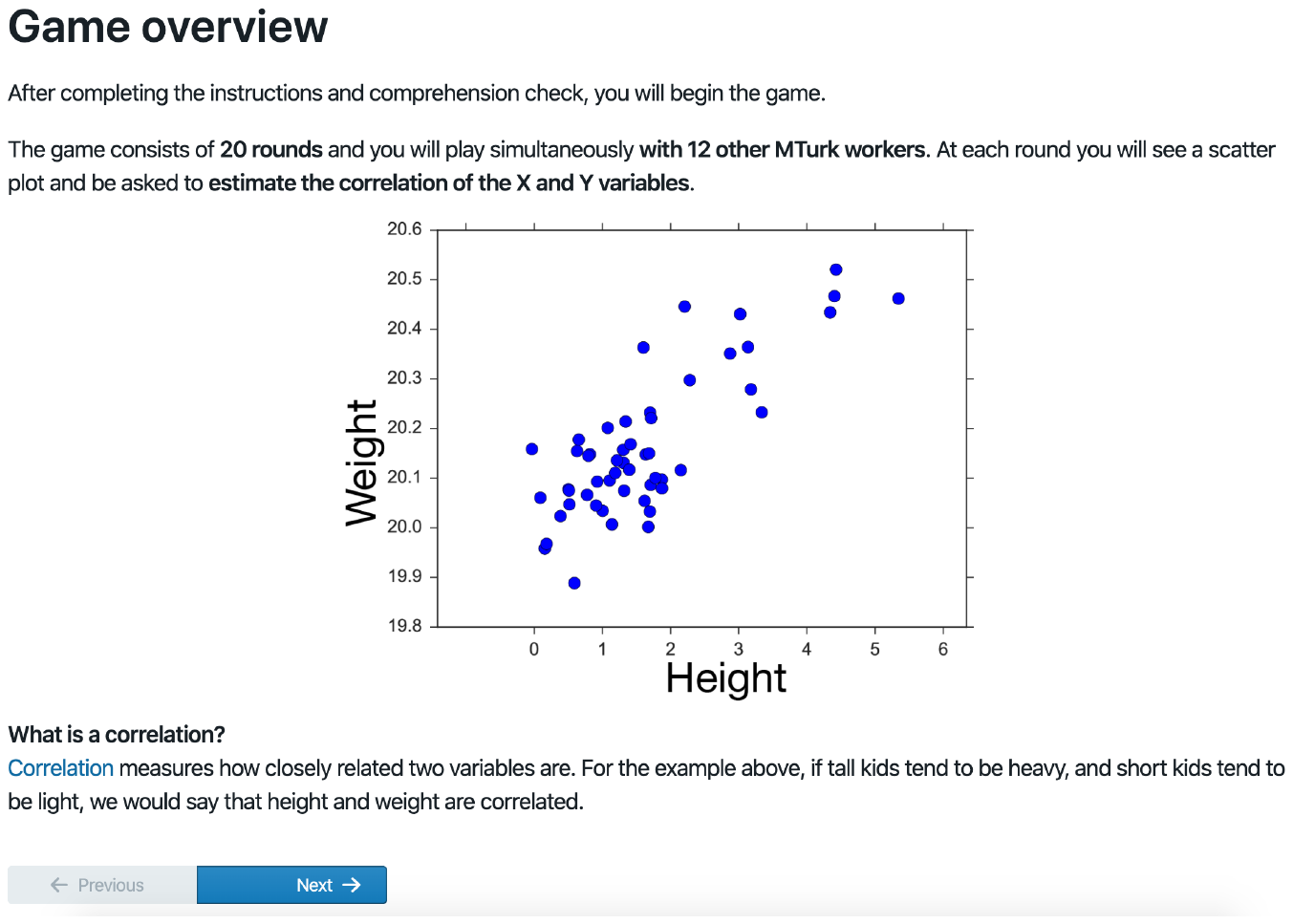}
\caption{Example of the instructions used in our experiments.}
\label{won:example1}
\end{figure}

\begin{figure}[H]
\includegraphics[width=1\textwidth]{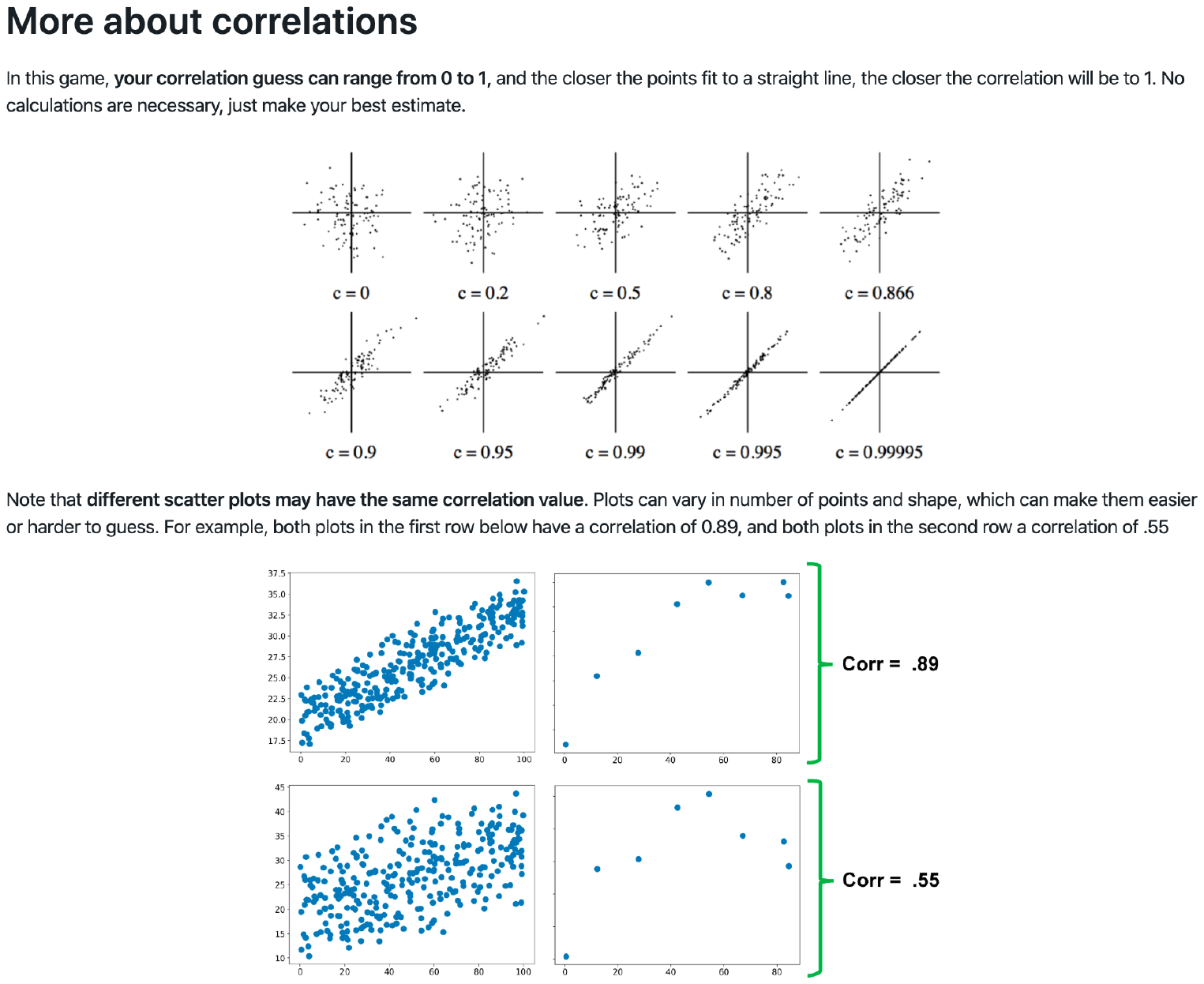}
\caption{Example of the instructions used in our experiments.}
\label{won:example2}
\end{figure}

\begin{figure}[H]
\centering
\includegraphics[width= .98 \textwidth]{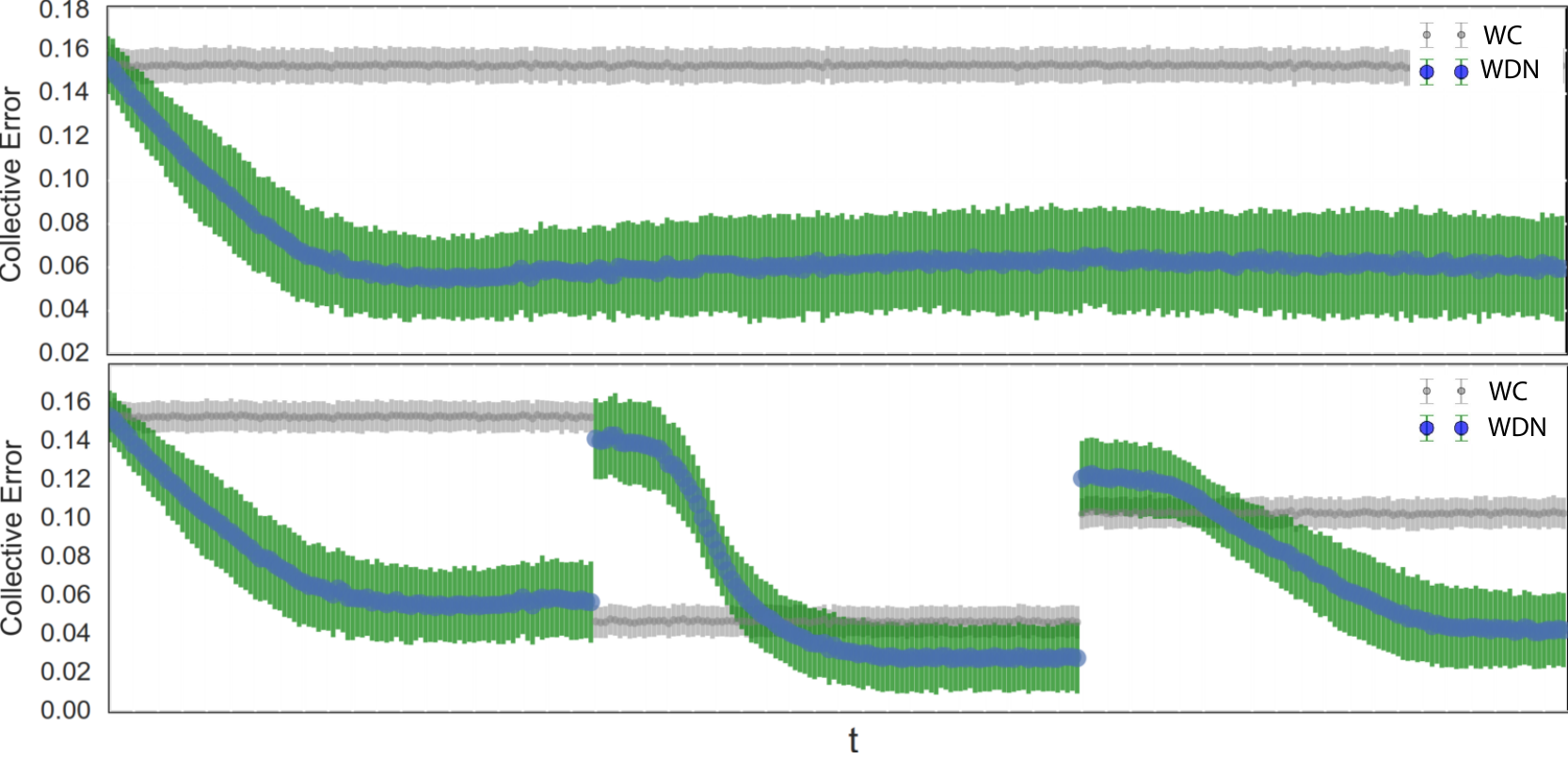}
\caption{Evolution of collective error: wisdom of the crowd (WC) and wisdom of the dynamic network (WDN). Panel A) stationary distribution of information among agents. Panel B) non-stationary information environment, shocks to the information distribution introduced at $t=\{100,200\}$}
\label{fig:won_woc_evolution}
\end{figure} 

\begin{figure}[H]
\centering
\includegraphics[width= 1\columnwidth]{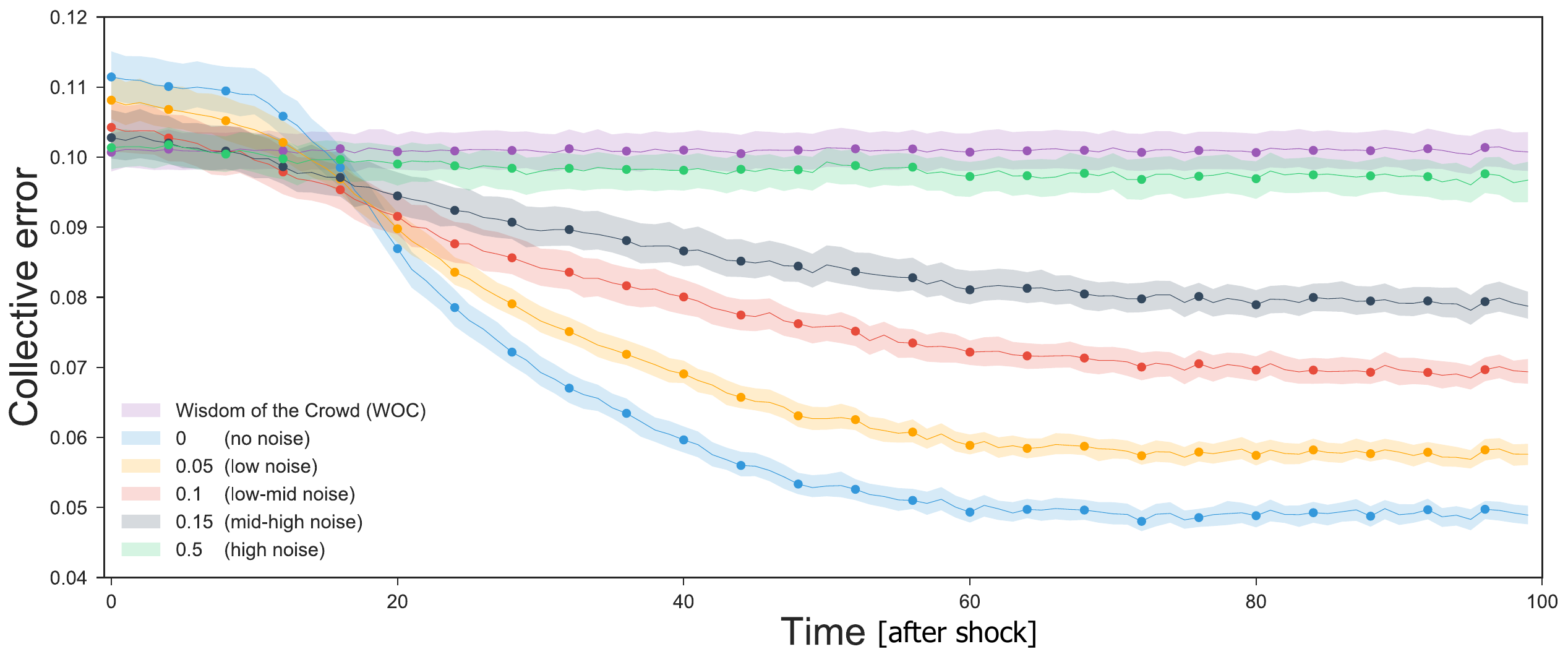}
\caption{As the nose level increases in the provided feedback, the collective performance degrades until it converges to the performance of the independent crowd.}
\label{fig:noise_level}
\end{figure}

\begin{figure}[H]
\centering
\includegraphics[width= 1\columnwidth]{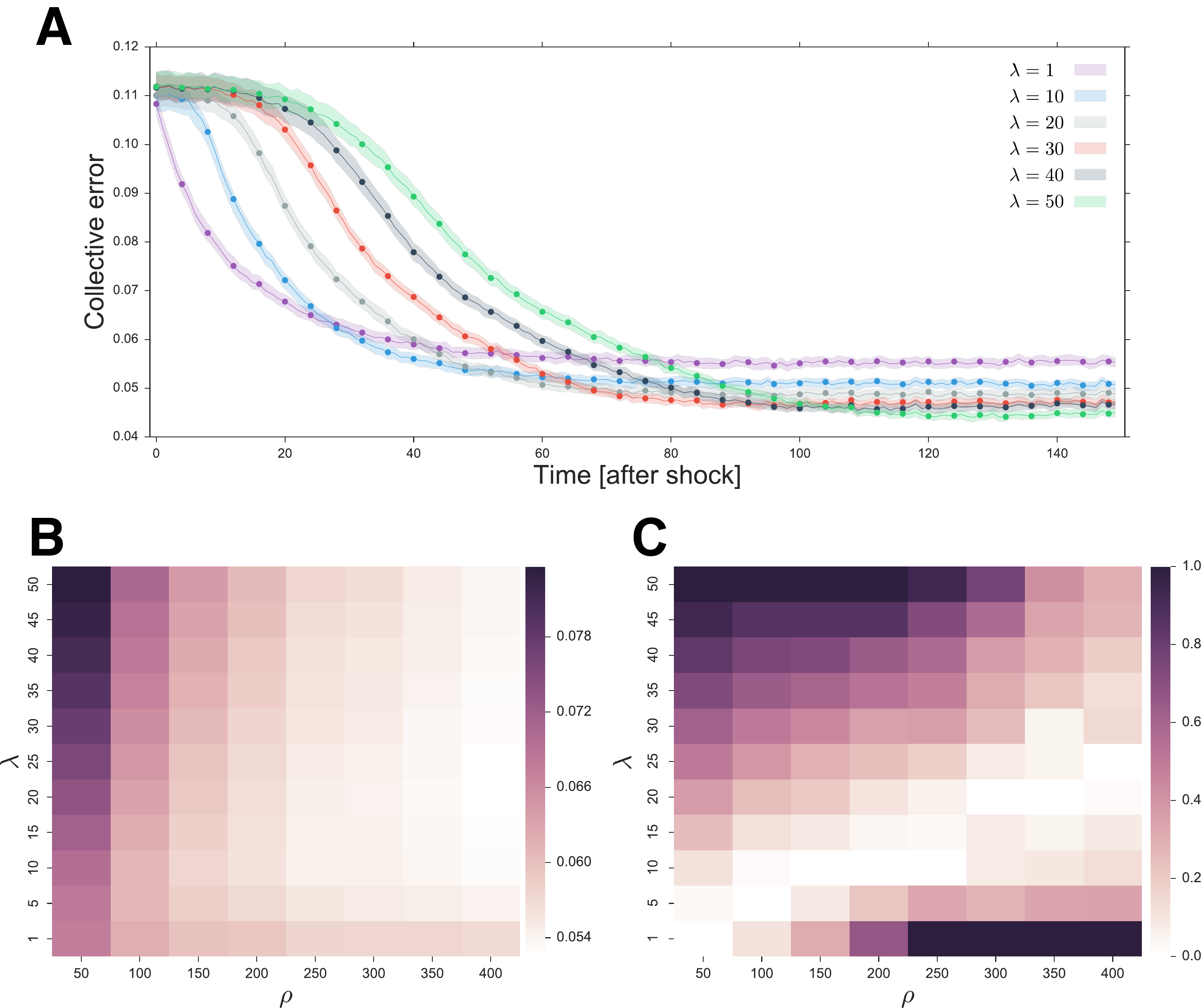}
\caption{Panel (\textbf{A}): Learning rates associated to different $\lambda$'s, where colored bands show $95\%$ confidence intervals. Panel (\textbf{B}): Effects of $\lambda$ and $\rho$ on collective error, where shades of orange indicate time-averaged collective error. Panel (\textbf{C}): Effects of $\lambda$ and $\rho$ on collective error, normalized per type of information environment ($\rho$ column).
}
\label{fig:lambda_rho}
\end{figure}

\section{Supplementary Tables}

\begin{table}[H]
    \label{tab:table1}
    \begin{center}
        \caption{Mixed effects models for individual and group errors for the first experiment (i.e., varying plasticity).}
        \begin{tabular}{|l|c|c|c|cc|}
            \hline
            \multicolumn{6}{|c|}{individual error $\sim$condition + (1 $|$ subject) + (1 $|$ group) + $\epsilon$}               \\ \hline
            \multicolumn{1}{|c|}{}   & $\beta$   & z-statistic & pvalue     & \multicolumn{2}{c|}{conf intervals} \\ \hline
            solo vs static: overall            & -0.055903 & -4.6239     & 3.7657e-06 & -0.079599        & -0.032207        \\
            solo vs static: adapted periods    & -0.056948 & -5.2265     & 1.7273e-07 & -0.078304        & -0.035592        \\
            static vs dynamic: overall         & -0.037745 & -4.6436     & 3.4239e-06 & -0.053676        & -0.021814        \\
            static vs dynamic: adapted periods & -0.053281 & -6.5555     & 5.5438e-11 & -0.069211        & -0.037351        \\ \hline
            \multicolumn{6}{|c|}{collective error $\sim$condition +  (1 | group) + $\epsilon$}                              \\ \hline
            solo vs static: overall            & -0.047132 & -8.4426     & 3.1031e-17 & -0.058073        & -0.03619         \\
            solo vs static: adapted periods    & -0.044902 & -6.6593     & 2.7506e-11 & -0.058118        & -0.031687        \\
            static vs dynamic: overall         & -0.030587 & -3.2548     & 0.0011349  & -0.049006        & -0.012168        \\
            static vs dynamic: adapted periods & -0.043012 & -4.4412     & 8.9455e-06 & -0.061994        & -0.02403         \\ \hline
        \end{tabular}
    \end{center}
\end{table}

\begin{table}[H]
    \label{tab:table2}
    \begin{center}
        \caption{Mixed effects models for individual and group errors for the second experiment (i.e., varying feedback).}
        \begin{tabular}{|l|c|c|c|cc|}
            \hline
            \multicolumn{6}{|c|}{individual error $\sim$condition + (1 $|$ subject) + (1 $|$ group) + $\epsilon$}               \\ \hline
            \multicolumn{1}{|c|}{}   & $\beta$   & z-statistic & pvalue     & \multicolumn{2}{c|}{conf intervals} \\ \hline
            solo vs  no feedback: overall&-0.0479 & -4.5405 & 5.612e-06& -0.0687& -0.0272\\
            solo vs  no feedback: adapted &-0.04160 & -4.1016 & 4.103e-05& -0.0614& -0.0217\\
             no feedback vs  self feedback: overall&-0.01450 & -1.3772 & 0.16846& -0.035& 0.00613\\
             no feedback vs  self feedback: adapted &-0.03677 & -3.5241 & 0.00042& -0.0572& -0.016\\
             self feedback vs  full feedback: overall&-0.032 & -3.2948 & 0.00098& -0.0510& -0.0129\\
             self feedback vs  full feedback:  adapted &-0.0414 & -4.9881 & 6.096e-07& -0.0577& -0.025\\ \hline
            \multicolumn{6}{|c|}{collective error $\sim$condition +  (1 $|$ group) + $\epsilon$}                              \\ \hline
            solo vs  no feedback: overall&-0.0325 & -4.908 & 9.1605e-07& -0.0456& -0.0195\\
            solo vs  no feedback: adapted &-0.0253 & -2.566 & 0.01026& -0.0446& -0.00598\\
             no feedback vs  self feedback: overall&-0.0186 & -1.665 & 0.0959& -0.04059& 0.00330\\
             no feedback vs  self feedback: adapted &-0.0383 & -2.935 & 0.0033& -0.0639& -0.0127\\
             self feedback vs  full feedback: overall&-0.021 & -2.252 & 0.024& -0.04066& -0.002\\
             self feedback vs  full feedback: adapted&-0.0299 & -2.991 & 0.0027& -0.0495& -0.0103\\ \hline
        \end{tabular}
    \end{center}
\end{table}

\begin{table}[H]
    \label{tab:table3}
    \begin{center}
        \caption{Mixed effects models for individual and group errors for the \textit{dynamic} condition in $E_1$ and \textit{full-feedback} condition in $E_2$.}
        \begin{tabular}{|l|c|c|c|cc|}
            \hline
            \multicolumn{6}{|c|}{individual error $\sim$condition + (1 $|$ subject) + (1 $|$ group) + $\epsilon$}               \\ \hline
            \multicolumn{1}{|c|}{}   & $\beta$   & z-statistic & pvalue     & \multicolumn{2}{c|}{conf intervals} \\ \hline
            dynamic $E_1$ vs  full feedback $E_2$: over all&-0.0025 & -0.263 & 0.792& -0.0214& 0.0163\\
            dynamic $E_1$ vs  full feedback $E_2$: adapted periods&0.0010 & 0.110 & 0.911& -0.017& 0.0197\\\hline
            \multicolumn{6}{|c|}{group error $\sim$condition + (1 | group) + $\epsilon$}               \\ \hline
            dynamic $E_1$ vs  full feedback $E_2$: over all&0.0009 & 0.090 & 0.927& -0.0187& 0.0205\\
            dynamic $E_1$ vs  full feedback $E_2$: adapted periods&0.004 & 0.411 & 0.680& -0.0152& 0.0234\\\hline
        \end{tabular}
    \end{center}
\end{table}

\begin{table}[H]
    \label{tab:table4}
    \begin{center}
        \caption{Mixed effects model for the network centralization as a function of being before the shock and after it (rounds $[11,15$) for the dynamic conditions with feedback. }
        \begin{tabular}{|l|c|c|c|cc|}
            \hline
            \multicolumn{6}{|c|}{freeman centralization $\sim$ after shock + (1 $|$ group) + $\epsilon$}               \\ \hline
            \multicolumn{1}{|c|}{}   & $\beta$   & z-statistic & p-value     & \multicolumn{2}{c|}{conf intervals} \\ \hline
            Intercept       &  0.580 &     23.652 &       0.000 &  0.532 &  0.628  \\
            After shock[T.True] & -0.046 &     -4.042 &       0.000 & -0.069 & -0.024  \\
            \hline
        \end{tabular}
    \end{center}
\end{table}

\end{document}